\documentclass{iopjournal}

\usepackage{subcaption}
\usepackage{amssymb, amsmath}
\usepackage{booktabs}
\usepackage[table]{xcolor}
\usepackage{multirow}
\usepackage{placeins}
\usepackage{textgreek}
\usepackage{cleveref}

\def\bracketbar{\smash{\hbox{\kern-6pt\raise1pt%
    \hbox{{\scalebox{.3}(}{\lower1.6pt\hbox{\bf--}}{\scalebox{.3})}}}}}
\providecommand{\varpm}{\mathbin{\vcenter{\hbox{%
  \oalign{\hfil$\scriptstyle+$\hfil\cr
          \noalign{\kern-.3ex}
          $\scriptscriptstyle({-})$\cr}%
}}}}
\providecommand{\varmp}{\mathbin{\vcenter{\hbox{%
  \oalign{$\scriptstyle({+})$\cr
          \noalign{\kern-.3ex}
          \hfil$\scriptscriptstyle-$\hfil\cr}%
}}}}
\usepackage{todonotes}

\begin{document}

\articletype{Paper} 

\title{Characterising the role of final state interactions on neutrino energy estimation in the DUNE and Hyper-K era}

\author{Stephen Dolan$^{1,*}$\orcid{0000-0002-2410-6550}, Jake McKean$^{2,*}$\orcid{0009-0005-6100-6195} and Laura Munteanu$^{1,*}$\orcid{0000-0002-2074-8898}}

\affil{$^1$CERN}

\affil{$^2$Kyoto University, Department of Physics, Kyoto, Japan}

\affil{$^*$Authors to whom any correspondence should be addressed.}

\email{stephen.joseph.dolan@cern.ch; mckean.jake.42u@st.kyoto-u.ac.jp; laura.munteanu@cern.ch}


\begin{abstract}
The Deep Underground Neutrino Experiment (DUNE) and Hyper-Kamiokande (Hyper-K) will measure neutrino oscillation parameters with an unprecedented precision that requires neutrino energy estimation to be controlled at the few-MeV level. A central challenge in achieving this is the modelling of the reinteractions of hadrons produced in neutrino-nucleus scatters with the residual nuclear medium, or final-state interactions (FSI). In this work we use state-of-the-art neutrino interaction event generators to review the impact of FSI modelling on the kinematic and calorimetric neutrino energy estimators used by Hyper-K and DUNE respectively, considering both the semi-classical intranuclear cascades (INCs) that dominate current simulations and a microscopic treatment based on a relativistic mean field calculation. We find that plausible variations of the FSI model introduce uncertainties on the neutrino energy estimation proxies that are at or above the precision on the energy scale control required for Hyper-K and DUNE projected neutrino oscillation sensitivities, highlighting the importance of careful FSI modelling to allow robust near detector constraints. We further demonstrate that the two experiments are sensitive to different aspects of the FSI models. Neutrino energy estimation at Hyper-K is most impacted by pion absorption and nuclear effects beyond the semi-classical paradigm, whilst the DUNE energy estimation is more affected by the modelling of how hadronic energy is shared between sources of visible and invisible energy in the detector. We discuss the implications of these findings for neutrino oscillation analyses and outline some of the key experimental and theoretical developments needed to bring FSI modelling uncertainties under control.
\end{abstract}
\section{Introduction}
\label{sec:intro}

The Deep Underground Neutrino Experiment (DUNE)~\cite{DUNE:2020ypp, DUNE:2020jqi} and Hyper-Kamiokande (Hyper-K)~\cite{Hyper-Kamiokande:2018ofw,Hyper-Kamiokande:2025fci} long-baseline (LBL) neutrino oscillation experiments, due to start operation within the coming few years, represent a leap forward in prospects for precision measurements of neutrino oscillations. Aiming to increase the statistics for neutrino oscillation analyses by more than an order of magnitude with respect to the currently-operating T2K~\cite{T2K:2011qtm} and NOvA~\cite{NOvA:2007rmc} LBL experiments, DUNE and Hyper-K are ideally placed: to make sub-percent level measurements of neutrino oscillation parameters; to probe CP violation in the lepton sector; to determine the neutrino mass ordering; and to search for physics beyond the three-flavour PMNS neutrino mixing paradigm~\cite{10.1143/PTP.28.870,Bilenky:1978nj,ParticleDataGroup:2024cfk,deBlas:2025gyz}. However, such precision will take neutrino oscillation measurements firmly out of the statistics limited era of current experiments into a regime where control over systematic uncertainties will underpin experimental success.

In general, LBL experiments operate by measuring the interactions of a predominantly muon neutrino beam twice: once at a near detector (ND) placed close to the neutrino beam production point, where oscillations are negligible, and again at a far detector (FD) placed downstream at a distance that approximately maximizes the neutrino oscillation probability. Experiments can operate using both neutrino and antineutrino enhanced beams. At the FD, experiments measure event rates ($N_{FD}$) of muon ($\mu$) or electron ($e$) neutrino ($\nu$) or antineutrino ($\bar\nu$) interactions that are sensitive to neutrino oscillation probabilities. Neglecting backgrounds and detector efficiency, these event rates can be written as: 

\begin{equation}
    N_{FD}^{\nu^{\bracketbar}_{\mu/e}} \propto P^{osc}_{\nu^{\bracketbar}_{\mu}\rightarrow\nu^{\bracketbar}_{\mu/e}} \: \Phi^{\nu^{\bracketbar}_{\mu}} \: \sigma^{\nu^{\bracketbar}_{\mu/e}},
    \label{eq:intosc}
\end{equation}

where $P^{osc}$ is the oscillation probability for muon (anti)neutrinos oscillating into either electron or muon (anti)neutrinos, $\Phi$ is the incoming muon (anti)neutrino flux and $\sigma$ is the interaction cross section for electron or muon (anti)neutrinos. The observed event rate at the ND is analogous but does not contain the oscillation probability term. Therefore, measurements at the ND are used to constrain $\sigma$ and $\Phi$.

Neutrino oscillation probabilities evolve as a function of the ratio of the ``baseline'' (i.e. the distance the neutrinos travel, a proxy for propagation time) and the neutrino energy ($E_\nu$). Whilst the baselines for LBL experiments are fixed, their neutrino fluxes are typically spread over one-to-several GeV and so the oscillation probability inferred in \autoref{eq:intosc} is the probability \textit{averaged} over the neutrino flux shape. However, a precise characterisation of neutrino oscillations requires inference of the neutrino oscillation probability as a function of $E_\nu$. 

A measurement of $N_{FD}(E_\nu)$ would directly characterise $P^{osc}(E_\nu)$ to the extent that $\sigma(E_\nu)$ and $\Phi(E_\nu)$ can be constrained. Unfortunately, in current and planned LBL experiments, the neutrino energy cannot be perfectly reconstructed or known a priori. Instead, experiments attempt to infer this by performing measurements of the final-state products of each neutrino interaction within the FD (here denoted as $\vec{\mathbf{x}}$), which significantly complicates oscillation measurements. Still neglecting backgrounds and detector effects, a differential event rate with respect to $\vec{\mathbf{x}}$ can be written as\footnote{Subscripts indicating the neutrino flavour and sign are dropped for legibility but would remain as in \autoref{eq:intosc}.}: 

\begin{equation}
  \frac{dN^{FD}}{d\vec{\mathbf{x}}} \propto \int{ dE_\nu \; \Phi(E_\nu) \; \sigma(E_\nu, \vec{\mathbf{x}}) \; M(\vec{\mathbf{x}} \mid E_\nu) \; P^{osc}(E_\nu)},
  \label{eq:diffosc}
\end{equation}

where $M(\vec{\mathbf{x}} \mid E_\nu)$ is a matrix mapping between $\vec{\mathbf{x}}$, the measured final state properties and the neutrino energy. It should be noted that the cross section itself also depends on $\vec{\mathbf{x}}$, further complicating the relationship between the differential event rate and the oscillation probability.  

From \autoref{eq:diffosc} it is clear that, even in this idealised case, a precise inference of the oscillation probability as a function of neutrino energy requires tight control of the incoming neutrino flux normalisations and shapes; the cross section as a function of the measured outgoing particle kinematics; and the mapping between the measured kinematics to the true neutrino energy. To avoid prematurely limiting the reach of DUNE and Hyper-K, which are expecting to collect $\mathcal{O}(10^4-10^5)$ events in their FD event samples~\cite{DUNE:2020jqi, Hyper-Kamiokande:2025fci}, each of these must be constrained at the few-percent level. 

In this work, we specifically characterise the physics that determines $M$ for the DUNE and Hyper-K primary neutrino energy estimation methods, building on previous work that has focussed on the impact of specific physics effects on one of the two~\cite{Liu:2025hpl,Ankowski:2014yfa,Ankowski:2015kya, Friedland:2018vry, Nagu:2019uco, DUNE:2021tad, Coyle:2025xjk}. We show that achieving the ultimate precision neutrino oscillation measurements targeted by the two experiments requires the neutrino energy estimation scale to be known at the few-MeV level. Afterwards, we show that \textit{final-state interactions} (FSI), which describe the reinteractions of hadrons produced in neutrino-nucleus scatters with the residual nucleus, introduce changes which impact the neutrino energy scale precision. We then quantify which aspects of FSI modelling which drive uncertainties in neutrino energy estimation for Hyper-K and DUNE, before contrasting the predictions of widely used semi-classical simulations with fully quantum-mechanical microscopic models. Finally, we discuss the implications of our results with the aim of informing how future dedicated measurements can constrain $M$ at the level required for next-generation LBL experiments.

\section{Analysis tools and methods}
\label{sec:methods}

\begin{figure}[h]
    \centering
    \begin{subfigure}[b]{0.45\linewidth}
      \centering
      \includegraphics[width=\linewidth,height=0.28\textheight,keepaspectratio]{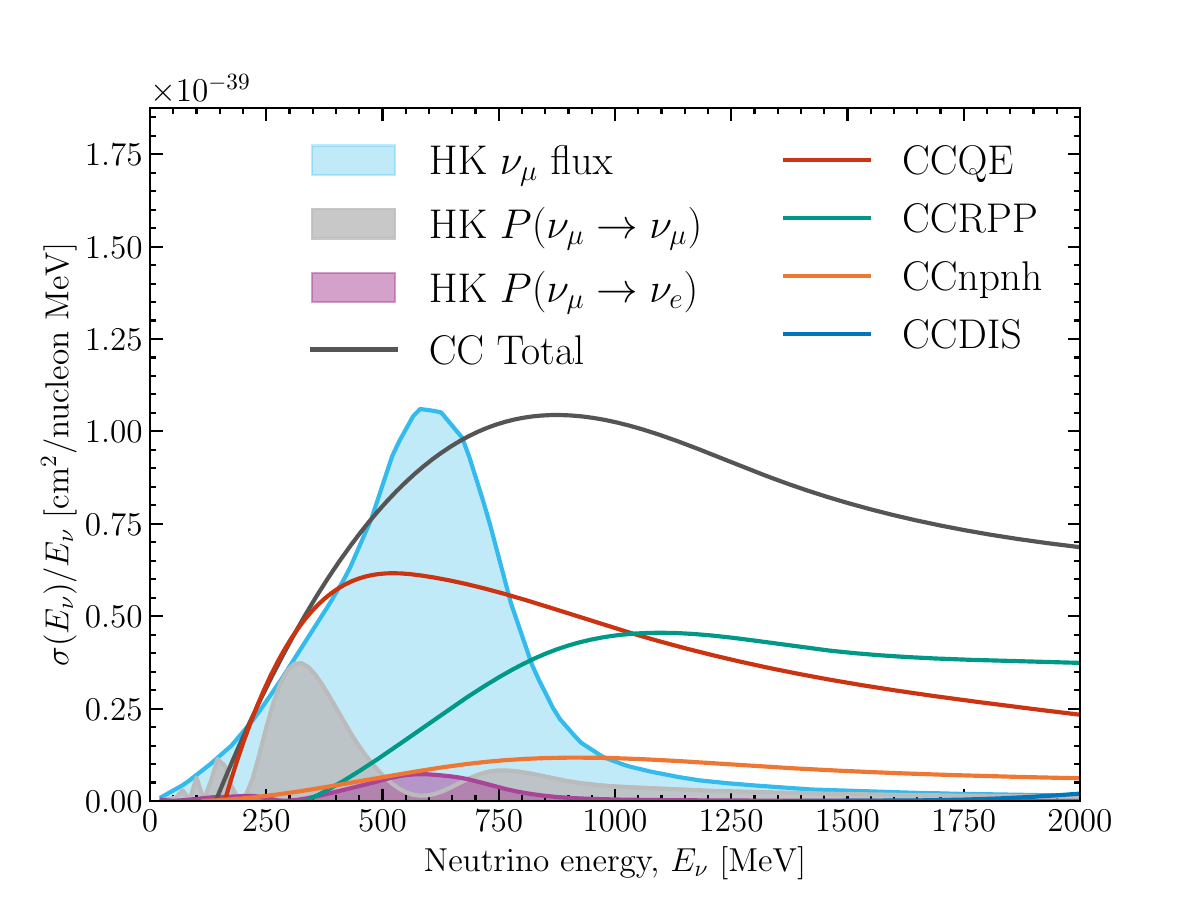}
      \caption{\footnotesize $\nu_\mu$ CC-H$_2$O, $\nu_\mu$ Hyper-K flux}
      \label{fig:sigmaEnu_HK_nu}
    \end{subfigure}
    \begin{subfigure}[b]{0.45\linewidth}
      \centering
      \includegraphics[width=\linewidth,height=0.28\textheight,keepaspectratio]{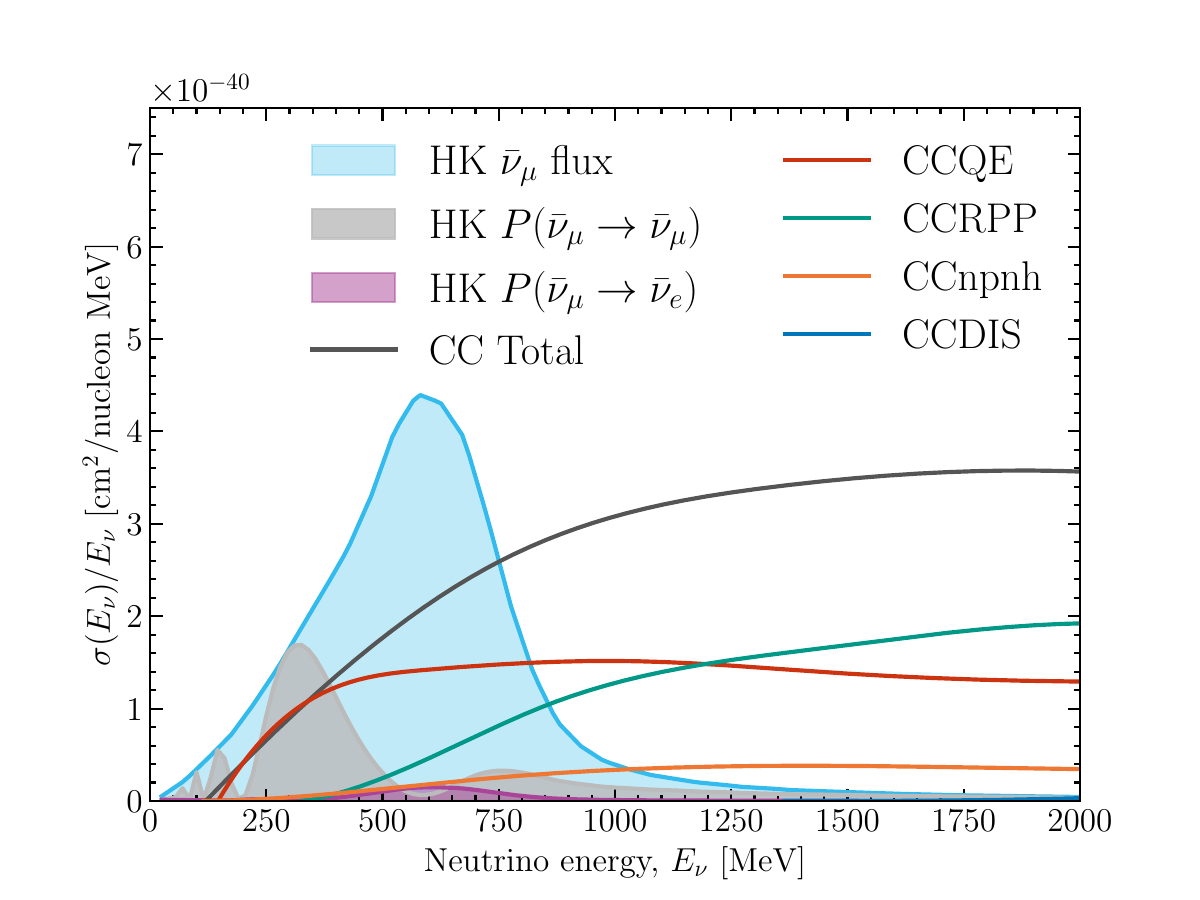}
      \caption{\footnotesize $\bar\nu_\mu$ CC-H$_2$O, $\bar\nu_\mu$ Hyper-K flux}
      \label{fig:sigmaEnu_HK_nubar}
    \end{subfigure}
    \begin{subfigure}[b]{0.45\linewidth}
      \centering
      \includegraphics[width=\linewidth,height=0.28\textheight,keepaspectratio]{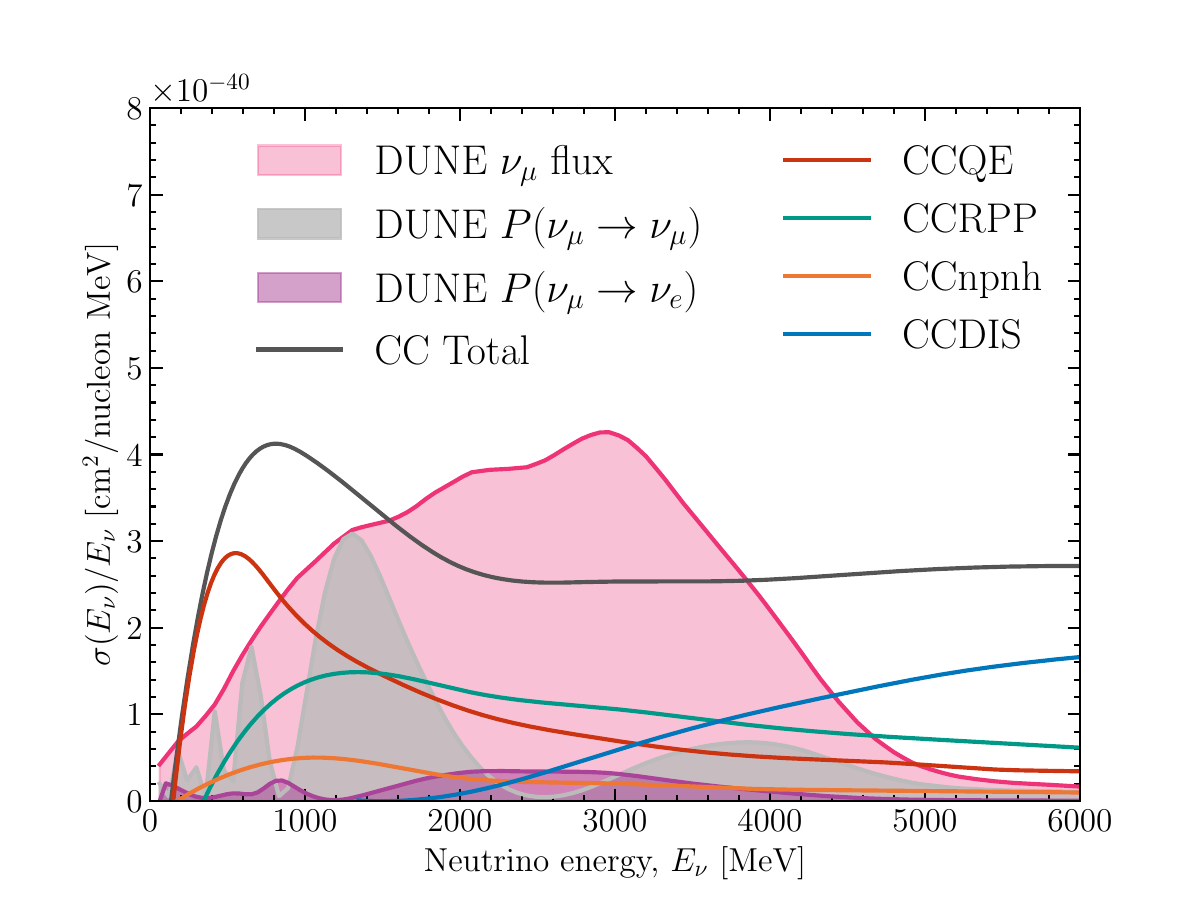}
      \caption{\footnotesize $\nu_\mu$ CC-Ar, $\nu_\mu$ DUNE flux}
      \label{fig:sigmaEnu_DUNE_nu}
    \end{subfigure}
    \begin{subfigure}[b]{0.45\linewidth}
      \centering
      \includegraphics[width=\linewidth,height=0.28\textheight,keepaspectratio]{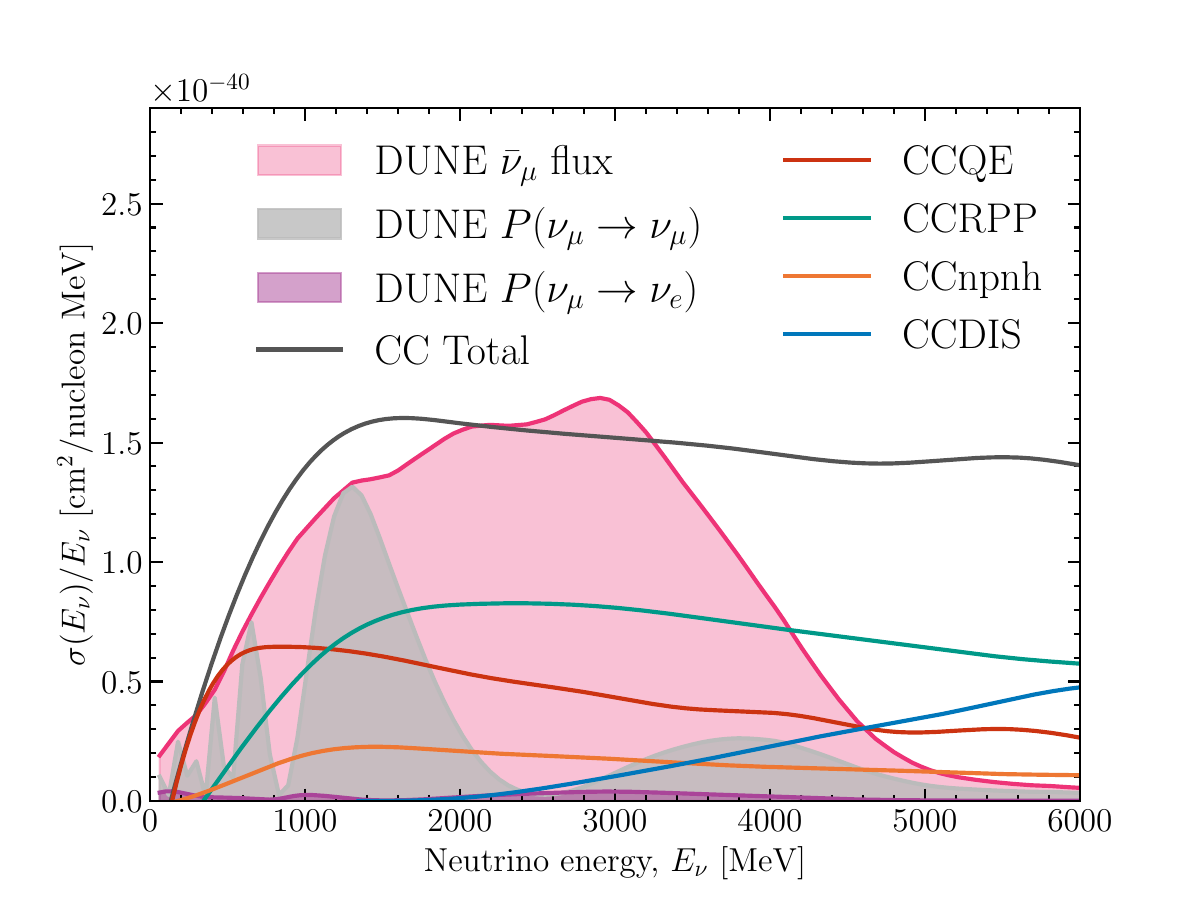}
      \caption{\footnotesize $\bar\nu_\mu$ CC-Ar, $\bar\nu_\mu$ DUNE flux}
      \label{fig:sigmaEnu_DUNE_nubar}
    \end{subfigure}
    \caption{The neutrino and antineutrino CC inclusive cross section on water and argon targets predicted by the NuWro simulation overlaid with the corresponding predicted neutrino fluxes from the Hyper-K and DUNE FD before and after considering neutrino oscillations (using the parameters given in \autoref{tab:OscProbParams}). The contribution to the cross section is broken down by neutrino interaction channel. Small contributions from coherent pion scattering are not shown.}
    \label{fig:sigmaEnu}
\end{figure}

In the few-GeV regime where DUNE and Hyper-K seek to measure neutrino oscillations, neutrinos primarily interact with nuclei. In this work, we simulate these interactions using state-of-the-art neutrino interaction Monte-Carlo event generators. We use the NuWro 25.03.1~\cite{Golan:2012wx,Golan2012nuwro,prasad2025}, GENIE v3.02.00~\cite{Andreopoulos:2009rq, Andreopoulos:2015wxa} and NEUT 6.1.3~\cite{Hayato:2002sd,Hayato:2009,Hayato:2021heg} event generators. These simulate each of the several contributing interaction channels at few-GeV neutrino energies~\cite{NuSTEC:2017hzk,Formaggio:2012cpf}. \autoref{fig:sigmaEnu} shows a comparison of the total cross section for different charged-current (CC) processes, as predicted by the NuWro. The four dominant interaction channels at DUNE and Hyper-K energies are: CC quasi-elastic (CCQE) interactions, in which a neutrino interacts with a bound nucleon within a nucleus and produces a charged lepton of the same flavour as the neutrino and a nucleon in the final state; multi-nucleon interactions, in which neutrinos interact with a correlated set of nucleons (CCnpnh, n-particle-n-hole interactions which are dominated by the CC2p2h contribution); CC resonant pion production (CCRPP), in which the struck nucleon is excited to a higher resonance (most often a $\Delta$(1232) baryon) which decays to usually produce a pion and a nucleon; and more inelastic interactions that begin to resolve nucleon quark structure which we refer to with the umbrella term CCDIS (for deep inelastic scattering, although exact definitions for this category vary~\cite{MINERvA:2025hzq}). For Hyper-K (\autoref{fig:sigmaEnu_HK_nu} and \autoref{fig:sigmaEnu_HK_nubar}), the dominant interaction channel is CCQE, but there are non-negligible contributions from CCnpnh and CCRPP. For DUNE (\autoref{fig:sigmaEnu_DUNE_nu} and \autoref{fig:sigmaEnu_DUNE_nubar}), CCQE, CCRPP and CCDIS all contribute significantly and similarly to the total cross section~\cite{DUNE:2020jqi}. 

\subsection{Modelling final state interactions with Monte Carlo generators}

Most event generators employ a factorised description of a neutrino interaction, where a neutrino interaction ``hard scatter'' is first simulated to produce hadrons within the target nucleus which are subsequently transported out, possibly re-interacting as they propagate, via a separate FSI routine. This approach uses semi-classical intranuclear cascade (INC) models to describe the FSI, which transport hadrons through the nucleus in discrete steps~\cite{Dytman:2021ohr,Salcedo:1987md,bertini}. An alternative approach, only recently implemented in generators, is a microscopic quantum mechanical calculation of cross section that considers FSI by calculating the outgoing hadronic wavefunctions in a nuclear potential~\cite{McKean:2025khb, gonzalez2019nuclear}. In this section, we will first detail these two different approaches to simulating FSI in neutrino generators, and then describe the generator configurations chosen for the rest of this work.

\subsubsection{Semi-classical intranuclear cascades}

INCs simulate FSI by transporting the outgoing hadrons through the residual nucleus using semi-classical methods. Each hadron is propagated as a classical particle whose interaction probability at every step is set by a local mean free path, computed from in-medium effective cross sections. At each interaction point, the hadron may scatter elastically, undergo charge exchange, be absorbed, or produce additional particles, with the choice of fate sampled probabilistically. Each additionally created particle is also propagated through the INC. By construction, the INC only redistributes energy and particle content among the final-state hadrons, leaving the cross section as a function of outgoing lepton kinematics for any interaction channel unchanged.

Most INC implementations used in neutrino event generators are ``space-like'' cascades, in which each hadron is propagated through the nucleus in discrete spatial steps, with the local interaction probability re-evaluated at each step. This is the approach used by both NEUT~\cite{Hayato:2021heg} and NuWro~\cite{Golan:2012wx}, as well as the hA2018 and hN2018 models in GENIE~\cite{Andreopoulos:2015wxa}. An alternative class of ``time-like'' cascades evolves the projectile and all spectator nucleons simultaneously in steps of space and time. The INCL++ model~\cite{Mancusi:2014eia,Boudard:2012wc}, available in GENIE, is an example of a time-like cascade and  is additionally able to describe processes that are often neglected in space-like frameworks, such as nuclear cluster emission. Once the cascade is complete, INCs can be coupled to a nuclear de-excitation routine (e.g. ABLA~\cite{Kelic:2009yg}) which dissipates the residual excitation energy of the remnant nucleus via the emission of additional particles, including nucleons or nuclear fragments. A comparison of many of these INC models can be found in Ref.~\cite{Dytman:2021ohr} and an additional review is provided in Ref.~\cite{Liu:2025hpl}. 

A well-known shortcoming of INC treatments is that they are applied incoherently on top of a hard-scatter cross section that is itself derived under generally different assumptions about the nuclear medium~\cite{Dytman:2021ohr}. This inconsistency between the initial interaction and the subsequent transport is a known source of theoretical uncertainty in any INC-based prediction. We note that there are some generators in which the treatment of FSI and the underlying nuclear model is fully consistent, such as GiBUU~\cite{Buss:2011mx} and Achilles~\cite{Isaacson:2022cwh}.

\subsubsection{Microscopic calculations}
Microscopic calculations of FSI aim to fix the issue of inconsistency that is present in INCs. In this case, the model calculates the hadron current from the bound- and scattered-state wavefunctions. The scattered-state wavefunctions are solutions to a wave equation including the nuclear potential, thus encoding the effect of the residual nuclear medium on the hadron dispersion relation. This is done in a consistent quantum mechanical and relativistic framework. This approach, when used with a real nuclear potential, encodes the case where the scattered hadron interacts with the residual nuclear medium and changes momentum and direction, but does not produce additional particles (coined ``elastic FSI'')~\cite{Nikolakopoulos:2022qkq}. When a complex nuclear potential is used, it also encodes the flux lost to inelastic processes and, in the CCQE case, constitutes an absolute lower bound of the cross section. 
However, this approach is not without drawbacks. Due to the complexity of calculating the scattered-state wavefunctions, it is computationally expensive. Additionally, effects such as additional nucleon creation, ejection and pion absorption are not simulated. Despite this, this approach has been implemented in the NEUT event generator via precomputed hadron tensor tables~\cite{McKean:2025khb}.
The approach is coupled with the NEUT INC to simulate both the elastic FSI from the microscopic calculation and the inelastic FSI from the NEUT INC. The NEUT INC has no inherent elastic FSI and, if one uses a real nuclear potential in the microscopic calculations, there is no danger of double-counting FSI effects.

\subsection{Event generator details}

As discussed, we consider simulations of neutrino interactions expected within DUNE and Hyper-K using the NuWro, GENIE and NEUT event generators. We use NuWro as our baseline simulation, GENIE to examine variations in simulations from a wide variety of INC models, and NEUT to consider microscopic FSI treatments.

We simulate $\nu_\mu$ and $\bar\nu_\mu$ interactions to be representative of those seen at DUNE and Hyper-K separately. For the DUNE case, we consider interactions using the experiment's $\nu_\mu$ and $\bar\nu_\mu$ fluxes~\cite{dunefluxurl} in neutrino-enhanced and antineutrino enhanced beam modes respectively. For DUNE we consider all simulated CC interactions, as this is the signal topology for the experiment's oscillation analysis~\cite{DUNE:2020jqi}. For the Hyper-K case, we generate analogous (anti)neutrino-water interactions using the T2K experiment's public flux~\cite{T2K:2012bge} (which uses the same neutrino beam as Hyper-K). For Hyper-K, we consider only CC interactions with no pions in the final state (CC0$\pi$ interactions) as the experiment's primary neutrino oscillation signal topology~\cite{Hyper-Kamiokande:2025fci}. Neutrino oscillations are applied via a reweighting of the input simulations using the \texttt{OscProb} framework~\cite{ProbOsc}, using the nominal parameters shown in \autoref{tab:OscProbParams}, taken from Ref.~\cite{ParticleDataGroup:2024cfk,ParticleDataGroup:2025neutrinomixing}\footnote{This parameter set corresponds to the NuFit 6.0 normal-ordering best fit without SK-atmospheric and IceCube-24 data~\cite{Esteban:2024eli}.} but with $\delta_{\mathrm{CP}}$ set to $-\pi/2$. All generator output is processed using NUISANCE~\cite{Stowell:2016jfr} into a common event format. In this work, we do not simulate contributions to DUNE and Hyper-K event rates due to backgrounds from topologies outside of the primary signal (e.g. the CC1$\pi$ background for Hyper-K or neutral current backgrounds) or from beam contamination (e.g. the $\nu_\mu$ contributions to antineutrino enhanced running mode). We also do not consider detector smearing or inefficiency. 

\begin{table}[h]
      \centering
      \begin{tabular}{cc}
      \hline \hline
         Parameter  & Value \\
         \hline
          $\theta_{12}$ [rad] & 0.58784 \\
          $\theta_{13}$ [rad] & 0.14870 \\
          $\theta_{23}$ [rad] & 0.84649 \\
          $\Delta m^{2}_{21}$ [eV$^{2}$] & $7.49 \times 10^{-5}$ \\
          $\Delta m^{2}_{32}$ [eV$^{2}$] & $2.459 \times 10^{-3}$ \\
          $\delta_{\mathrm{CP}}$ & $-\pi/2$ \\
          Mass Ordering & Normal Ordering \\
           \hline \hline
      \end{tabular}
      \caption{Oscillation parameters used in the simulations, taken from Table 14.7 in the PDG~2025 review of neutrino masses, mixing and oscillations (Ref.~\cite{ParticleDataGroup:2024cfk,ParticleDataGroup:2025neutrinomixing}), corresponding to the NuFit 6.0 normal-ordering best fit without SK-atmospheric and
  IceCube-24 data~\cite{Esteban:2024eli}; $\delta_{\mathrm{CP}}$ is fixed to $-\pi/2$ for the purposes of this work.}
      \label{tab:OscProbParams}
  \end{table}

\subsubsection{NuWro}

CCQE interactions are simulated using a spectral function approach based on Refs.~\cite{benhar1994spectral, Ankowski:2014yfa} for water and on Refs.~\cite{PhysRevD.105.112002, Banerjee_2024, Ankowski:2025umq} for argon. Two- and three-body current (CC2p2h and CC3p3h respectively)
interactions are simulated with the model from the Valencia group~\cite{PhysRevC.102.024601, PhysRevD.111.036032:nuwro_npnh}.
Pion production below the deep inelastic scattering (DIS) region is simulated with the Hybrid
model from the Ghent group~\cite{Yan:2024kkg}. DIS generation uses the GRV98 parton distribution functions with Bodek-Yang corrections~\cite{Gluck:1998xa,Bodek:2002vp}
together with PYTHIA~\cite{Sjostrand:2006za} for hadronisation at high hadronic invariant mass ($W>1.9$~GeV$/c^2$). For intermediate hadronic masses, $1.6<W<1.9$~GeV$/c^2$, NuWro implements a linear transition between the Hybrid and DIS models.
FSI are simulated using NuWro's custom INC~\cite{Golan:2012wx}, a space-like
cascade that uses the Salcedo-Oset model~\cite{Salcedo:1987md} for low-energy
pion-nucleus interactions and free $NN$ cross sections, with in-medium
corrections, for nucleon reinteractions. 

\subsubsection{GENIE}

In addition to NuWro, we also use simulations from the GENIE generator, provided in Refs.~\cite{CWretNuisTutNuInt2024,IoPReview}, to study the effect of more substantial changes to the INC model. GENIE is a widely-used neutrino generator (notably in LAr-based experiments) and has a broad range of available models and tunes. GENIE is the only neutrino generator to provide four different INC models which can be combined with different hard scatter models. 

To describe primary interactions, we use the
\texttt{G18\_10a\_00\_000} \texttt{GENIE} configuration~\cite{GENIE:2021zuu},
which uses a local Fermi gas nuclear ground state and the model from the Valencia group
for CCQE and CCnpnh interactions~\cite{Nieves:2011pp, gran2013neutrino, Schwehr:2016pvn}, although for CCnpnh it uses an older version of the model than is in NuWro and only considers the CC2p2h channel. For CCRPP interactions, GENIE uses the Berger-Sehgal pion production models~\cite{Berger:2007rq,PhysRevD.77.059901}, and for DIS it uses the GRV98 parton distribution functions with Bodek-Yang corrections~\cite{Gluck:1998xa,Bodek:2002vp}. Hadronisation is modelled either with PYTHIA~\cite{SJOSTRAND199474,Sjostrand:2006za} (at invariant masses, $W>3.0$~GeV/$c^2$), the custom AGKY model~\cite{Yang:2009zx} ($W<2.3$~GeV/$c^2$), or an interpolation between
them~\cite{GENIE:2021wox}. Four samples are generated with this hard-scatter configuration held fixed, each using a different INC available within GENIE: the default hA2018 model~\cite{GENIE:2021zuu}, the hN2018 model~\cite{GENIE:2021zuu}, the Li\`ege Intranuclear Cascade model (INCL)~\cite{Boudard:2012wc}, and the GEANT4 Bertini Cascade model (G4BC)~\cite{Heikkinen:2003sc}, the latter accessed through GENIE's dedicated interface~\cite{genie_g4bc}. These cascades differ substantially in the physics choices they make. Very broadly, hA2018 and hN2018 are widely-used models that are tuned extensively to hadron scattering data, while INCL and G4BC are more sophisticated hadron transport models with a more careful treatment of the nuclear medium and a richer set of final-state channels. Further details and comparisons of these FSI models are available in Refs.~\cite{Dytman:2021ohr,ershova2022study,Liu:2025hpl}. Holding the hard-scatter model fixed across the four samples ensures that any variation between them is attributable to the INC alone. We note that the hA2018 and hN2018 models share much of their pion FSI treatment with the NEUT and NuWro cascades. 

\subsubsection{ED-RMF in NEUT}
\label{subsec:edrmf_model}

We use the NEUT event generator in order to access its implementation of microscopic FSI model calculations for CCQE interactions, recently implemented and detailed in Ref.~\cite{McKean:2025khb}. In this model, bound- and final-state wavefunctions are used to define the hadron current, which is then used to describe the hadron tensor as a sum of single particle relativistic mean field (RMF) states. The final-state wavefunctions are solutions to the Dirac equation in the presence of a relativistic nuclear potential. Doing this encodes elastic FSI into the final-state particle information which is then used to seed the NEUT INC and perform inelastic FSI. It has been shown that using real relativistic nuclear potentials poses little risk of double counting FSI effects with such INCs~\cite{Nikolakopoulos:2022qkq}\footnote{It is important to note that this is only true if the event generator does not have a dedicated elastic rescattering calculation in the INC, or it is turned off, as is the case for NEUT.}. The nuclear potentials shown in this work are the energy-dependent RMF (ED-RMF)~\cite{Gonzalez_Jimenez_2020, PhysRevC.105.025502} and relativistic plane-wave impulse approximation (RPWIA). The RPWIA model considers no nuclear potential for the scattered nucleon, and so is a useful reference model without a microscopic FSI treatment. The ED-RMF potential is constructed by applying a softening function to the energy-independent RMF potential of the bound-state system~\cite{Gonzalez-Jimenez:2019ejf}. The softening function is a function of scattered nucleon kinetic energy ($T_{N}$) and at low values of $T_{N}$ it is close to one; this ensures that the initial and final states are solutions to the same Dirac equation and encodes natural Pauli blocking. This model, due to its computational complexity, is currently only implemented for CCQE interactions.

\section{Neutrino energy estimation}
\label{sec:enurec}

LBL experiments usually use one of two classes of methods for neutrino energy estimation. This choice is tailored to their detector technology and the ranges of neutrino energies they cover. 

\subsection{Kinematic neutrino energy estimation}

The kinematic method of neutrino energy estimation infers neutrino energy from the kinematics of outgoing particles under some set of approximations. Its most common application derives an estimator for neutrino energy, $E^{\mathrm{QE}}_{\nu}$, by assuming that an incoming neutrino undergoes a CCQE interaction with a single neutron at rest within the nucleus which is subject to some fixed nuclear binding energy, $E_{\mathrm{b}}$:

\begin{equation}
    E^{\mathrm{QE}}_{\nu} = \frac{m_p^2-m_\ell^2-(m_n-E_{\mathrm{b}})^2+2E_\ell(m_n-E_{\mathrm{b}})}{2(m_n-E_{\mathrm{b}}-E_\ell+p_\ell^z)}.
    \label{eq:enuqe}
\end{equation}

Here $m_{p/\ell/n}$ is the mass of a proton, outgoing lepton or neutron; and $E_\ell$ and $p_\ell^z$ are the outgoing lepton energy and projected lepton momentum along the direction of the incoming neutrino, respectively. $E_{\mathrm{b}}$ is the mean nuclear binding energy and is set to 27~MeV in this work. For antineutrino interactions the proton and neutron masses are swapped. This form of kinematic energy estimation only requires the reconstruction of the outgoing lepton kinematics, making it applicable in water Cherenkov detectors where outgoing hadrons may be below the Cherenkov threshold. Since it is built assuming a CCQE interaction, it is usually only applied to subsamples of identified interactions that contain a high proportion of CCQE events and so is best suited for experiments with neutrino beams with energies predominantly below the point at which other channels become dominant ($\lesssim 1$~GeV). In the context of T2K and Hyper-K, which share a neutrino beam peaked at $\sim$0.6~GeV, \autoref{eq:enuqe} is applied to a selection of CC0$\pi$ interactions\footnote{Whilst not considered in this work, within T2K a modified version of the~\autoref{eq:enuqe} formula is applied to a subsample of interactions rich in pion production by replacing the proton mass by the $\Delta(1232)$ mass~\cite{T2K:2021xwb}.} which are predicted by the NuWro simulation to contain 78.7\% (85.1\%) of $\nu_\mu$ ($\bar\nu_\mu$) CCQE interactions and make up 61.0\% (72.1\%) of the total CC event rate considering the (anti)neutrino enhanced flux. 

\subsection{Calorimetric neutrino energy estimation}

Experiments with an FD capable of low-threshold particle tracking and calorimetry can use alternative metrics to estimate neutrino energy, based on summing energy deposits within their detectors. This applies to the Liquid Argon TPC of the DUNE experiment as well as the scintillator detectors of the NOvA experiment, which are both sensitive to the energy that charged particles leave inside the detector down to negligible threshold. Neglecting very rare events with heavy mesons and baryons, a common variable used to estimate the energy recorded by calorimetrically summing energy deposits is:

\begin{equation}
  E^{\mathrm{avail}}_{\nu} = E_{\ell} + \sum_{i=p,\pi^{\pm}}{T_i} + \sum_{i=\pi^{0}, \gamma,e^{\pm}}{E_i}.
  \label{eq:enuavail}
\end{equation}

Here, $E^{\mathrm{avail}}_{\nu}$ is the estimated neutrino energy approximated as the outgoing lepton energy plus the sum of the kinetic energies of protons and charged pions and the sum of the total energy of neutral pions and photons. This variable was first introduced in Ref.~\cite{MINERvA:2015ydy}. Only the kinetic energy of protons is considered as their mass energy is already present in the initial state of the interaction\footnote{The effects of the 1.3 MeV proton-neutron mass difference and nuclear binding energy are neglected here}. The same applies to light nuclear fragments produced in the interaction (deuterons, tritons and $\alpha$ particles), which contribute only their kinetic energy. Neutrons are assumed to not deposit any energy within the detector that can be associated with the interaction. The total energy of neutral pions is included because their kinetic and mass energy is converted to photons in a prompt decay and the latter subsequently deposit almost their total energy in the detector following electron-positron pair production. Charged pions undergo frequent secondary interactions and partially decay to neutrinos. \autoref{eq:enuavail} makes a conservative assumption that none of the pions' mass energy is visible. Very small contributions from heavier nuclear remnants, hyperons and any other heavy mesons and baryons are neglected in the formula for this work. 

At the other extreme, an alternative  neutrino energy estimator could be written as:

\begin{equation}
  E^{\mathrm{had}}_{\nu} = E_{\ell} + \sum_{i=p}{T_i} + \sum_{i=\pi^{\pm}, \pi^{0}, \gamma,e^{\pm}}{E_i},
  \label{eq:enuhad}
\end{equation}

where the total charged pion energy is considered. As in $E^{\mathrm{avail}}_{\nu}$, protons and light nuclear fragments enter only through their kinetic energy and heavier hadrons are neglected. It is likely that the true visible energy deposit within a DUNE TPC lies somewhere between the two. The two estimators are contrasted and further discussed in Ref.~\cite{Wilkinson:2022dyx}.

\subsection{Neutrino energy estimation and precision oscillation measurements}
\label{subsec:osc_spectra}

\begin{figure}[htb]
\centering
  \includegraphics[width=\linewidth]{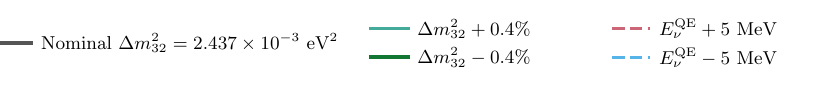}\\[2pt]
  \begin{subfigure}[b]{0.40\linewidth}
    \centering
    \includegraphics[width=\linewidth]{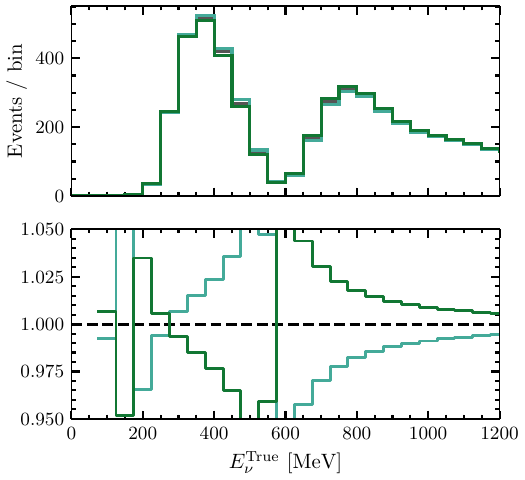}
    \caption{\footnotesize$E^{\mathrm{true}}_{\nu}$, CC0$\pi$-H$_2$O, Hyper-K $\nu_\mu$ osc.\ flux}
    \label{fig:Enuspect_dm32_HK_true}
  \end{subfigure}
  \begin{subfigure}[b]{0.40\linewidth}
    \centering
    \includegraphics[width=\linewidth]{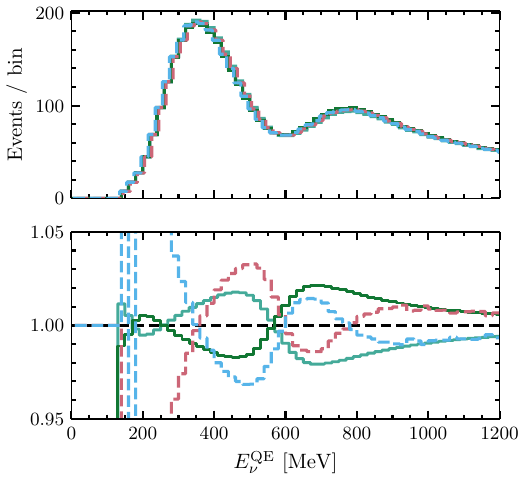}
    \caption{\footnotesize$E^{\mathrm{QE}}_{\nu}$, CC0$\pi$-H$_2$O, Hyper-K $\nu_\mu$ osc.\ flux}
    \label{fig:Enuspect_dm32_HK_reco}
  \end{subfigure}\\[8pt]
  \includegraphics[width=\linewidth]{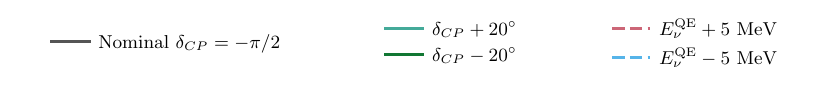}\\[2pt]
  \begin{subfigure}[b]{0.40\linewidth}
    \centering
    \includegraphics[width=\linewidth]{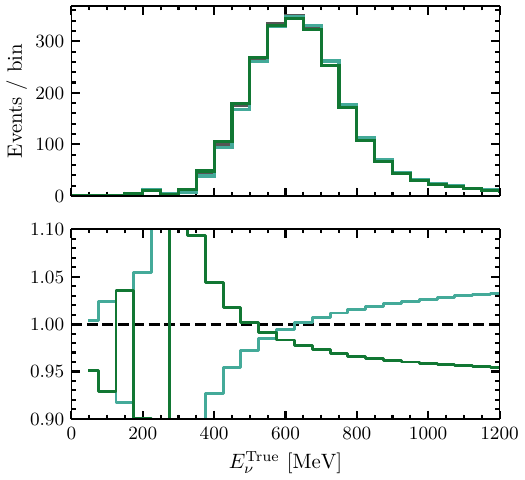}
    \caption{\footnotesize$E^{\mathrm{true}}_{\nu}$, CC0$\pi$-H$_2$O, Hyper-K $\nu_e$ osc.\ flux}
    \label{fig:Enuspect_dcp_HK_true}
  \end{subfigure}
  \begin{subfigure}[b]{0.40\linewidth}
    \centering
    \includegraphics[width=\linewidth]{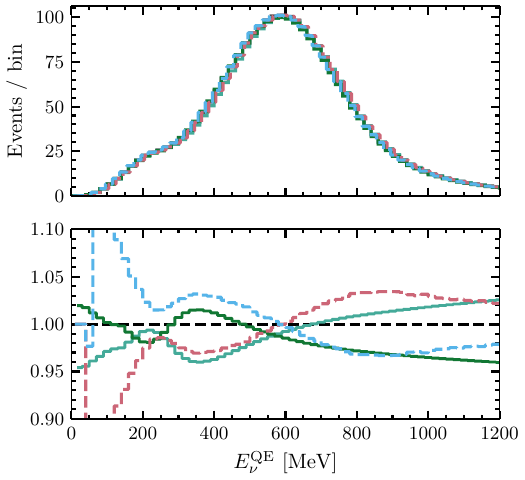}
    \caption{\footnotesize$E^{\mathrm{QE}}_{\nu}$, CC0$\pi$-H$_2$O, Hyper-K $\nu_e$ osc.\ flux}
    \label{fig:Enuspect_dcp_HK_reco}
  \end{subfigure}
  \caption{The NuWro simulated rate of CC0$\pi$ $\nu_\mu$ or $\nu_e$ interactions on water at the Hyper-K FD using the oscillated Hyper-K flux (see \autoref{tab:OscProbParams}) as a function of true ($E^{\mathrm{true}}_{\nu}$) or estimated ($E_{\nu}^{\mathrm{QE}}$) neutrino energy. For the former, variations of oscillation parameters corresponding to the Hyper-K ultimate target precision are shown. For the latter, these are compared to $E_{\nu}^{\mathrm{QE}}$ biased by $\pm$5~MeV.}
  \label{fig:Enuspect_HK}
\end{figure}

\begin{figure}[tbp]
\centering
  \includegraphics[width=\linewidth]{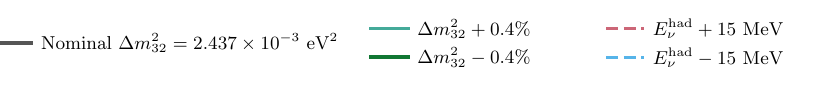}\\[2pt]
  \begin{subfigure}[b]{0.40\linewidth}
    \centering
    \includegraphics[width=\linewidth]{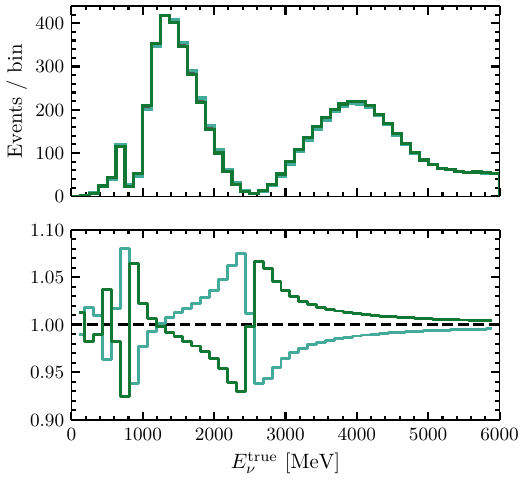}
    \caption{\footnotesize$E^{\mathrm{true}}_{\nu}$, CCINC-Ar, DUNE $\nu_\mu$ osc.\ flux}
    \label{fig:Enuspect_DUNE_dm32_true}
  \end{subfigure}
  \begin{subfigure}[b]{0.40\linewidth}
    \centering
    \includegraphics[width=\linewidth]{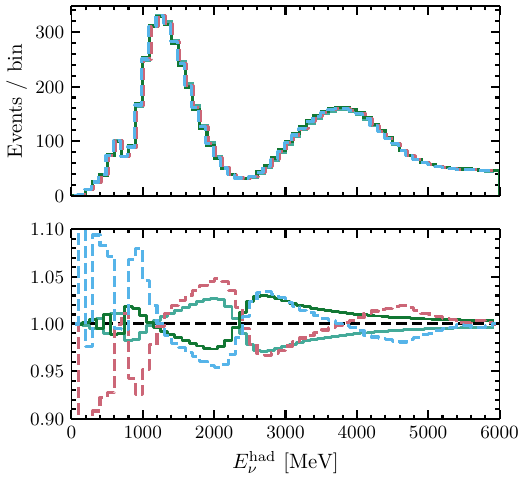}
    \caption{\footnotesize$E^{\mathrm{had}}_{\nu}$, CCINC-Ar, DUNE $\nu_\mu$ osc.\ flux}
    \label{fig:Enuspect_DUNE_dm32_reco}
  \end{subfigure}\\[8pt]
  \includegraphics[width=\linewidth]{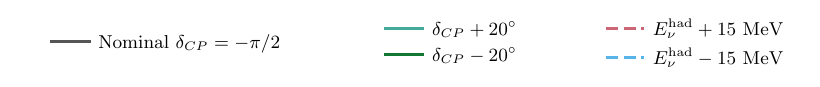}\\[2pt]
  \begin{subfigure}[b]{0.40\linewidth}
    \centering
    \includegraphics[width=\linewidth]{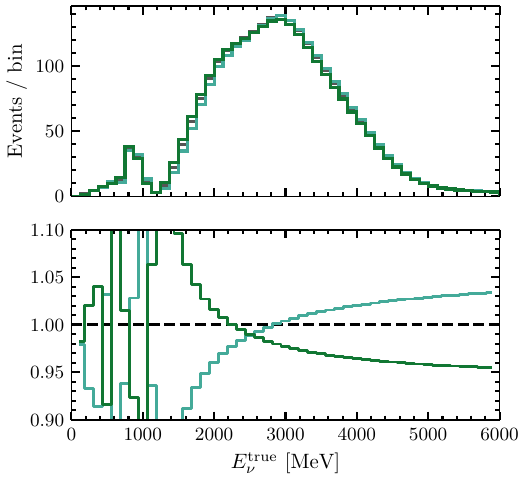}
    \caption{\footnotesize$E^{\mathrm{true}}_{\nu}$, CCINC-Ar, DUNE $\nu_e$ osc.\ flux}
    \label{fig:Enuspect_DUNE_dcp_true}
  \end{subfigure}
  \begin{subfigure}[b]{0.40\linewidth}
    \centering
    \includegraphics[width=\linewidth]{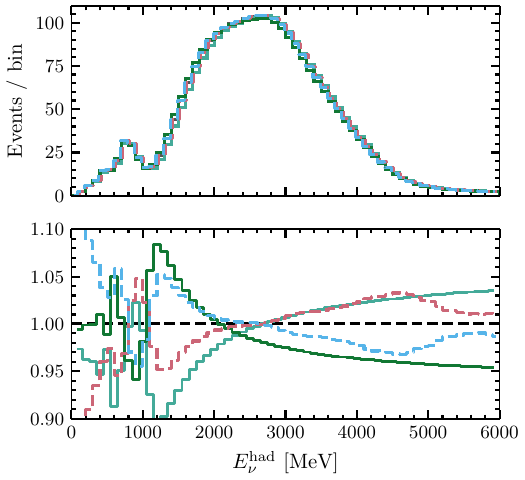}
    \caption{\footnotesize$E^{\mathrm{had}}_{\nu}$, CCINC-Ar, DUNE $\nu_e$ osc.\ flux}
    \label{fig:Enuspect_DUNE_dcp_reco}
  \end{subfigure}
  \caption{The NuWro simulated rate of CC inclusive $\nu_\mu$ or $\nu_e$ interactions on argon at the DUNE FD using the oscillated DUNE flux (see \autoref{tab:OscProbParams}) as a function of true ($E^{\mathrm{true}}_{\nu}$) or estimated ($E^{\mathrm{had}}_{\nu}$) neutrino energy. For the former, variations of oscillation parameters corresponding to the DUNE ultimate target precision are shown. For the latter, these are compared to $E^{\mathrm{had}}_{\nu}$ biased by $\pm$15~MeV.}
  \label{fig:Enuspect_DUNE}
\end{figure}

Precise neutrino energy estimation is needed to accurately infer neutrino oscillation parameters from the shape of the oscillated neutrino spectra seen in LBL experiments. Within the three-flavour PMNS neutrino mixing framework, this is especially relevant for measurements of $\Delta m_{32}^2$ and $\delta_{\mathrm{CP}}$, particularly when the latter is close to values of $\pm\pi/2$ (implying maximal CP violation). Modifications to these parameters have a significant impact on the oscillated spectra shape~\cite{Hyper-Kamiokande:2025fci,DUNE:2020jqi}. To demonstrate this, \autoref{fig:Enuspect_dm32_HK_true}, \autoref{fig:Enuspect_dcp_HK_true} and \autoref{fig:Enuspect_DUNE_dm32_true}, \autoref{fig:Enuspect_DUNE_dcp_true} show the projected event rate of muon ($N_{FD}^{\nu_{\mu}}$) or electron neutrino ($N_{FD}^{\nu_{e}}$) CC0$\pi$ interactions as a function of \textit{true} neutrino energy using the Hyper-K and DUNE fluxes respectively. The integrated event rates have been scaled to approximately 10 years of expected data taking according to the numbers provided in Refs.~\cite{DUNE:2020jqi,Hyper-Kamiokande:2025fci}, which amounts to a total $\nu_\mu$ event rate of 13436 and 8845, and an oscillated $\nu_e$ rate of 2591 and 2475 for DUNE and Hyper-K respectively. NC backgrounds, wrong-sign contaminations and intrinsic $\nu_e$ contributions are ignored. In these figures, the baseline NuWro simulation has been reweighted to consider small variations of either $\Delta m_{32}^2$ ($\pm 0.4\%$) or $\delta_{\mathrm{CP}}$ ($\pm 20^{\circ}$) around the nominal neutrino oscillation probability considered, which closely corresponds to the experiments' target precisions~\cite{Hyper-Kamiokande:2025fci, DUNE:2020jqi}. This illustrates that the position of the oscillation maximum in the FD muon neutrino spectra is sensitive to variations of $\Delta m_{32}^2$, whilst the peak position of the electron neutrino appearance spectra is sensitive to changes in $\delta_{\mathrm{CP}}$. This stems directly from the neutrino oscillation probability. The leading-order disappearance probability:
\begin{equation}
P(\nu_\mu \to \nu_\mu) \simeq 1 - \sin^2 2\theta_{23}\,
\sin^2\!\left(\frac{\Delta m^2_{32}\, L}{4 E_\nu}\right),
\end{equation}
depends on $\Delta m^2_{32}$ only through the combination $\Delta m^2_{32}$ and $L / E_\nu$, so changing $\Delta m^2_{32}$ shifts the position of oscillation dip along the energy axis. The shift is larger at low $E_\nu$ and reverses sign on either side of the dip, so $\Delta m^2_{32}$ is read off from the position of the suppression and cannot be absorbed by an overall change in normalisation. In the appearance channel, the leading order $\delta_{\mathrm{CP}}$-dependent interference term of the oscillation probability can be written compactly as:
\begin{equation}
\Delta P_{\delta_{\mathrm{CP}}}(\nu_\mu \to \nu_e) \propto
\cos\!\left(\frac{\Delta m^2_{32}\, L}{4 E_\nu} + \delta_{\mathrm{CP}}\right),
\label{eq:osc_dcp}
\end{equation}
so $\delta_{\mathrm{CP}}$ enters as a phase shift of an argument which depends on $1/E_\nu$. The observable response to $\delta_{\mathrm{CP}}$ is the change in oscillation probability which manifests as a change in the observed event rate. This change corresponds to the derivative of \autoref{eq:osc_dcp}:
\begin{equation}
\frac{\partial\, \Delta P_{\delta_{\mathrm{CP}}}}{\partial \delta_{\mathrm{CP}}} \propto
-\sin\!\left(\frac{\Delta m^2_{32}\, L}{4 E_\nu} + \delta_{\mathrm{CP}}\right),
\end{equation}
and whether a small change in $\delta_{\mathrm{CP}}$ resembles a rate change or a shape change is decided by the symmetry of this function about the first oscillation maximum, where $\Delta m^2_{32}\, L / 4 E_\nu = \pi/2$. At $\delta_{\mathrm{CP}} = 0$ or $\pi$ the derivative is $\mp\sin(\Delta m^2_{32}\, L / 4 E_\nu)$, which is symmetric about the maximum and largest there: a shift in $\delta_{\mathrm{CP}}$ moves the event rate at the oscillation peak and changes both sidebands in the same direction, mimicking a normalisation change. At $\delta_{\mathrm{CP}} = \pm\pi/2$ (i.e. maximal) the derivative is $\mp\cos(\Delta m^2_{32}\, L / 4 E_\nu)$, which is antisymmetric about the energy of the oscillation peak and vanishes there. A shift in $\delta_{\mathrm{CP}}$ leaves the peak rate unchanged and instead moves the probability along the neutrino energy axis. This distortion is therefore largest precisely when CP violation is maximal, but it is identical for $+\pi/2$ and $-\pi/2$, so determining the sign of $\delta_{\mathrm{CP}}$ requires input from the antineutrino spectrum, in which the $\sin\delta_{\mathrm{CP}}$ term enters with the opposite sign. Implications of oscillation-parameter and energy-scale variations on the antineutrino oscillation spectra and on the ratios of $\nu_e$ to $\bar\nu_e$ appearance events are discussed in \autoref{app:anu}.

\autoref{fig:Enuspect_dm32_HK_reco}, \autoref{fig:Enuspect_dcp_HK_reco} and \autoref{fig:Enuspect_DUNE_dm32_reco}, \autoref{fig:Enuspect_DUNE_dcp_reco} show the event rate as a function of a neutrino energy estimator ($E^{\mathrm{QE}}_{\nu}$ and $E^{\mathrm{had}}_{\nu}$ for the Hyper-K and DUNE cases, respectively) rather than true neutrino energy. This demonstrates that $\Delta m_{32}^2$ and $\delta_{\mathrm{CP}}$ affect the shape of the FD muon and electron \textit{estimated} neutrino energy spectra similarly to the \textit{true} one. The figures also show alternative estimated energy spectra in which the nominal oscillation parameters are used but the neutrino energy estimators are biased by a constant 5~MeV for the Hyper-K case or 15~MeV for the DUNE case. These numbers were chosen empirically, but they broadly correspond to a shift of 0.5\% of the means of DUNE and Hyper-K energies. It is clear that the difference between the nominal and alternative reconstructed energy spectra is comparable to the difference between the nominal and altered values of $\Delta m_{32}^2$ and $\delta_{\mathrm{CP}}$, giving an approximate scale for the precision with which the neutrino energy scale must be controlled. Overall, this illustrates the importance of constraining neutrino energy estimation at the 5-15~MeV level for next-generation LBL experiments to meet their ultimate precision goals\footnote{Whilst the focus of this work concerns how FSI modelling might introduce neutrino energy scale uncertainties, it is the overall energy scale uncertainty that must be kept to the 5-15~MeV level. This can have contributions from other nuclear effects or from the detector smearing.}.

\FloatBarrier

\section{The impact of FSI on neutrino energy estimation}
\label{sec:results}

In this section, we describe how FSI affects the neutrino energy estimators introduced in \autoref{sec:enurec}. We first characterise their performance in the absence of FSI, then show how FSI modifies it, and finally isolate specific FSI processes to quantify their impact. The performance of the estimators is set by the hadronic system and the estimator definition, and is essentially independent of the outgoing lepton flavour; we therefore use muon (anti)neutrinos throughout, averaging over the oscillated $\nu_\mu$ ($\bar\nu_\mu$) flux for the neutrino(antineutrino) case, with the $\nu_e$/$\bar\nu_e$ results being very similar. We discuss the implications of the muon-versus-electron oscillated spectra choices in \autoref{sec:discussion}.

\subsection{Neutrino energy estimation bias without FSI}

\begin{figure*}[tbh]
 \centering
   \begin{subfigure}[b]{0.40\linewidth}
     \centering
     \includegraphics[width=\linewidth]{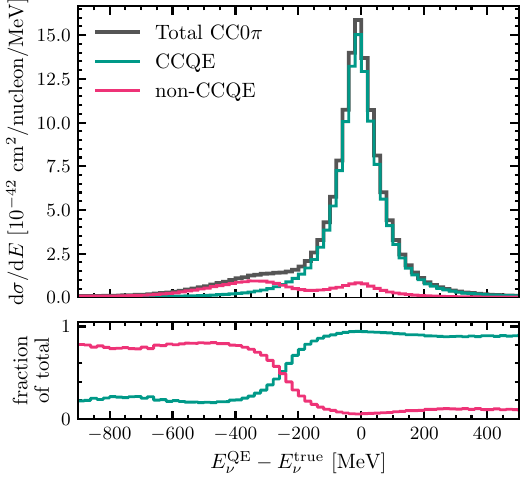}        \caption{\footnotesize$E^{\mathrm{QE}}_{\nu}$, CC0$\pi$-H$_2$O, Hyper-K $\nu_\mu$ osc. flux}
     \label{fig:noFSIEnuQEnu}
   \end{subfigure}
   \begin{subfigure}[b]{0.40\linewidth}
     \centering
     \includegraphics[width=\linewidth]{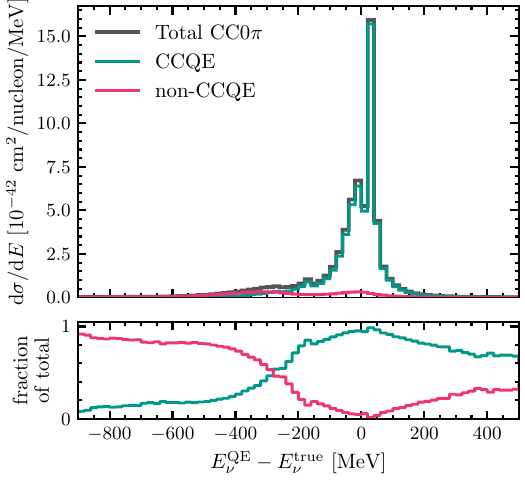}       \caption{\footnotesize$E^{\mathrm{QE}}_{\nu}$, CC0$\pi$-H$_2$O, Hyper-K $\bar\nu_\mu$ osc. flux}
     \label{fig:noFSIEnuQEanu}
   \end{subfigure}

 \caption{The NuWro simulated neutrino energy estimation bias (using $E^{\mathrm{QE}}_{\nu}$) for CC0$\pi$ $\nu_\mu$ or $\bar\nu_\mu$ interactions on a water target at the Hyper-K FD using the oscillated Hyper-K flux (see \autoref{tab:OscProbParams}), split by contributions from CCQE and non-CCQE interactions.}
\label{fig:noFSI_HK}
\end{figure*}

\begin{figure*}[tbh]
 \centering

  \begin{subfigure}[b]{0.40\linewidth}
    \centering
    \includegraphics[width=\linewidth]{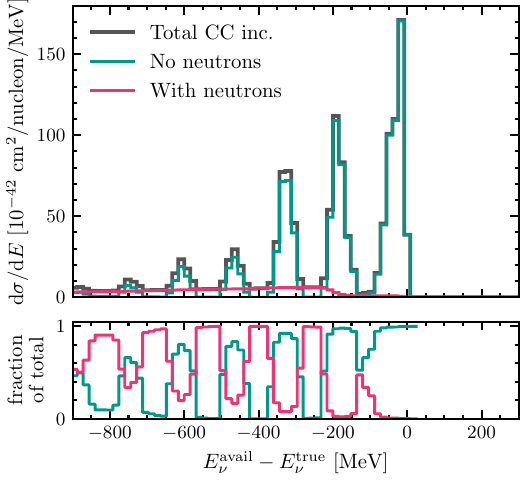}    \caption{\footnotesize$E^{\mathrm{avail}}_{\nu}$, CCINC-Ar, DUNE $\nu_\mu$ osc. flux}
    \label{fig:noFSIEnuavailnu}
  \end{subfigure}
  \begin{subfigure}[b]{0.40\linewidth}
    \centering
    \includegraphics[width=\linewidth]{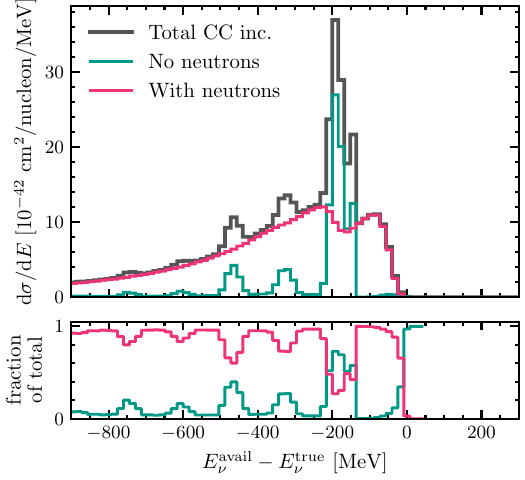}
    
    \caption{\footnotesize$E^{\mathrm{avail}}_{\nu}$, CCINC-Ar, DUNE $\bar\nu_\mu$ osc. flux}
    \label{fig:noFSIEnuavailanu}
    
  \end{subfigure}
  \begin{subfigure}[b]{0.40\linewidth}
    \centering
    \includegraphics[width=\linewidth]{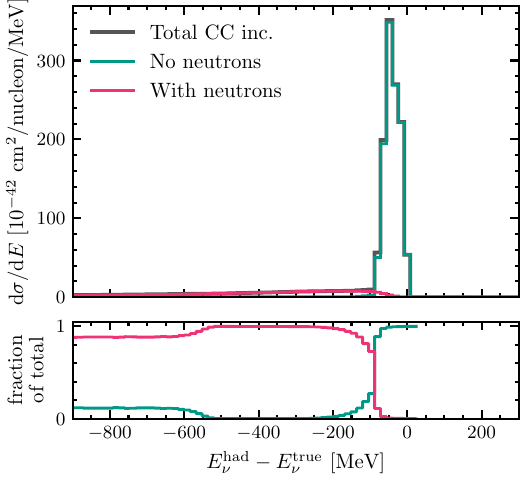}      
    \caption{\footnotesize$E^{\mathrm{had}}_{\nu}$, CCINC-Ar, DUNE $\nu_\mu$ osc. flux}
    \label{fig:noFSIEnuhadnu}
  \end{subfigure}
  \begin{subfigure}[b]{0.40\linewidth}
    \centering
    \includegraphics[width=\linewidth]{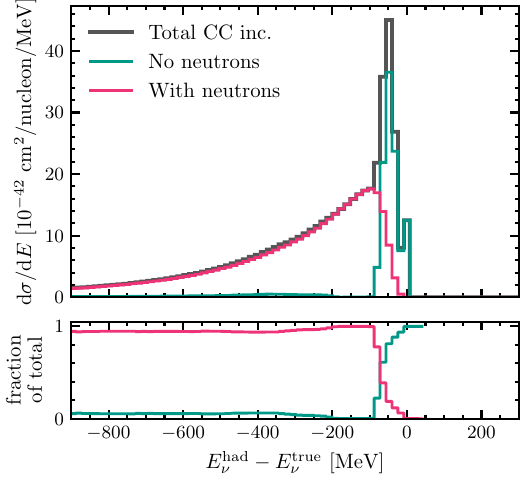}        \caption{\footnotesize$E^{\mathrm{had}}_{\nu}$, CCINC-Ar, DUNE $\bar\nu_\mu$ osc. flux}
    \label{fig:noFSIEnuhadanu}
  \end{subfigure}

 \caption{The NuWro simulated neutrino energy estimation bias (using $E^{\mathrm{had}}_{\nu}$ or $E^{\mathrm{avail}}_{\nu}$) for CC inclusive $\nu_\mu$ or $\bar\nu_\mu$ interactions on an argon target at the DUNE FD using the oscillated DUNE flux (see \autoref{tab:OscProbParams}), split by contributions from interactions that do and do not produce final state neutrons.}
\label{fig:noFSI_DUNE}
\end{figure*}

\autoref{fig:noFSI_HK} and \autoref{fig:noFSI_DUNE} show the difference between true neutrino energy and the neutrino energy estimators introduced in \autoref{sec:enurec} ($E^{\mathrm{QE}}_{\nu}$ for Hyper-K, $E^{\mathrm{avail}}_{\nu}$ and $E^{\mathrm{had}}_{\nu}$ for DUNE), henceforth referred to as the energy estimation \textit{bias}, using NuWro simulations without any FSI for the Hyper-K and DUNE case respectively. These and subsequent distributions are shown as differential cross sections with respect to the energy estimation bias, which is proportional to the probability of finding an interaction with a particular bias. The distributions are shown separately for neutrino and antineutrino interactions, broken down by contributions that help characterise the physics responsible for the imperfect energy estimation even before FSI.  

The bias in the neutrino energy estimation directly stems from the approximations made in building the estimators. In the case of $E^{\mathrm{QE}}_{\nu}$, \autoref{eq:enuqe} is built assuming a CCQE interaction on a stationary nucleon with some fixed binding energy. Most of the time, none of these conditions are entirely met, causing the bias shown in \autoref{fig:noFSI_HK}. Firstly, nucleons in the nucleus are not stationary, they move with some \textit{Fermi motion} which is isotropic and distributed between 0 and $\sim$230 MeV/c~\cite{JLabE91013:2003gdp}. This motion causes interactions to have their energy misestimated by the momentum of the struck nucleon before the interaction takes place and is responsible for the majority of the spread seen for CCQE interactions. Secondly, nucleons inside the nucleus do not have a single fixed nuclear binding energy but rather have a distribution of possible values (see e.g. Ref.\cite{benhar1994spectral}). This causes a further small spread (at a maximum of the few 10s of MeV scale of binding energies) and a potential non-zero mean bias if the average real binding energy is not the $E_b$ assumed in \autoref{eq:enuqe}. Thirdly, it is not possible to isolate a pure sample of CCQE events. Even for CC0$\pi$ interactions with no FSI, interactions with bound states of multiple nucleons (CCnpnh interactions) leave final states that can be indistinguishable from CCQE interactions. Since \autoref{eq:enuqe} assumes an interaction with a single target nucleon, the energy of CCnpnh interactions is estimated incorrectly. The bias seen is similar for neutrino and antineutrino interactions, with the main differences driven by the slightly changing fraction of CCQE to CCnpnh interactions. The small bump in \autoref{fig:noFSIEnuQEanu} at $\sim$-200~MeV is due to $\Lambda^0$ producing interactions. 

There is one case where the $E^{\mathrm{QE}}_{\nu}$ formula is almost exact: when an antineutrino interaction happens on a hydrogen atom (which is the case for slightly less than 2/18 of the interactions on a water target), which is not subject to nuclear effects. In this case $E^{\mathrm{QE}}_{\nu}$ estimates the neutrino energy perfectly except for an offset related to the chosen value of the binding energy in \autoref{eq:enuqe} (as hydrogen should have zero binding energy). This contribution is clearly visible in \autoref{fig:noFSIEnuQEanu} as a large peak centred at a value corresponding to the difference between \autoref{eq:enuqe} evaluated with $E_b=0$ MeV and the chosen value of 27~MeV.

In contrast, $E^{\mathrm{avail}}_{\nu}$ and $E^{\mathrm{had}}_{\nu}$ assume that all charged particle energy deposited in a detector is identified and so the bias observed in \autoref{fig:noFSI_DUNE} is from unaccounted energy from other sources. The dominant contribution is from energy that goes unseen as it is lost to final-state neutrons. This is typically a much larger fraction of the total interaction energy for antineutrino interactions than for neutrino interactions (this is clearly shown by comparing \autoref{fig:noFSIEnuavailnu} and \autoref{fig:noFSIEnuhadnu} to \autoref{fig:noFSIEnuavailanu} and \autoref{fig:noFSIEnuhadanu}). Beyond neutrons, energy lost to overcoming nuclear binding energy is also unseen, causing biases at the level of 10-50 MeV, which dominates the no neutron contribution to Figures \ref{fig:noFSIEnuhadnu} and \ref{fig:noFSIEnuhadanu}. For Figures  \ref{fig:noFSIEnuavailnu} and \ref{fig:noFSIEnuavailanu} the mass energy of charged pions is also assumed to go unseen. This causes additional bias in discrete units of pion masses, seen very clearly for neutrino interactions. This additional bias also affects antineutrino interactions, but discrete peaks in the bias are not so obvious as their impact is small relative to the continuous bias driven by energy lost to neutrons. The very small contributions to the no neutron population at large negative bias in Figures \ref{fig:noFSIEnuhadnu} and \ref{fig:noFSIEnuhadanu} come from interactions that produce hyperons, causing a bias of at least their mass, as well as other heavy mesons and baryons that are not explicitly accounted for in \autoref{eq:enuavail} and \autoref{eq:enuhad}. 

\subsection{The impact of INC FSI}
\label{subsec:impactofINC}

\begin{figure}[tbh]
\centering
  \begin{subfigure}[b]{0.40\linewidth}
    \centering
    \includegraphics[width=\linewidth]{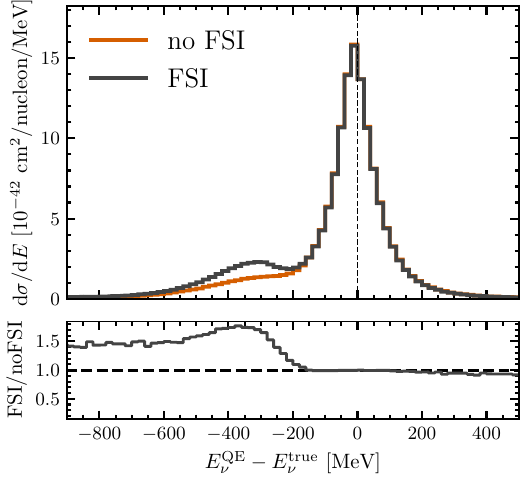}    \caption{\footnotesize$E^{\mathrm{QE}}_{\nu}$, CC0$\pi$-H$_2$O, Hyper-K $\nu_\mu$ osc. flux}
    \label{fig:piabs_nu}
  \end{subfigure}
  \begin{subfigure}[b]{0.40\linewidth}
    \centering
    \includegraphics[width=\linewidth]{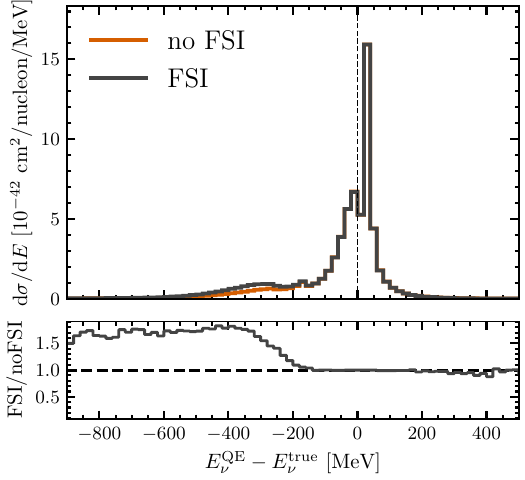}    \caption{\footnotesize$E^{\mathrm{QE}}_{\nu}$, CC0$\pi$-H$_2$O, Hyper-K $\bar\nu_\mu$ osc. flux}
    \label{fig:piabs_anu}
  \end{subfigure}

 \caption{The NuWro simulated neutrino energy estimation bias (using $E^{\mathrm{QE}}_{\nu}$) for CC0$\pi$ $\nu_\mu$ or $\bar\nu_\mu$ interactions on a water target at the Hyper-K FD using the oscillated Hyper-K flux (see \autoref{tab:OscProbParams}), split for cases when the NuWro INC is and is not applied.}
\label{fig:piabs}
\end{figure}

\begin{figure}[tbh]
 \centering
   \begin{subfigure}[b]{0.40\linewidth}
    \centering
    \includegraphics[width=\linewidth]{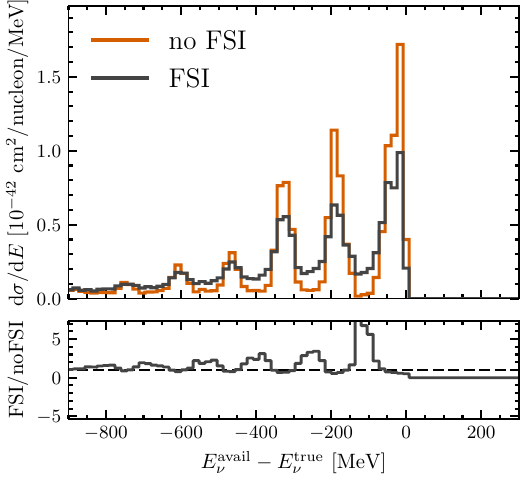}    \caption{\footnotesize$E^{\mathrm{avail}}_{\nu}$, CCINC-Ar, DUNE $\nu_\mu$ osc. flux}
    \label{fig:EreccalEavailnu}
  \end{subfigure}
  \begin{subfigure}[b]{0.40\linewidth}
    \centering
    \includegraphics[width=\linewidth]{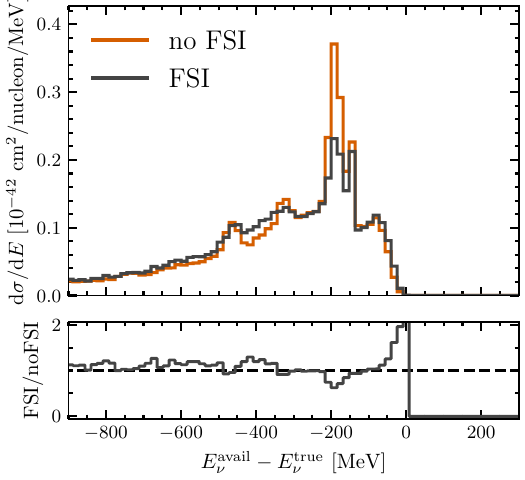}    
    \caption{\footnotesize$E^{\mathrm{avail}}_{\nu}$, CCINC-Ar, DUNE $\bar\nu_\mu$ osc. flux}
    \label{fig:EreccalEhadnu}
  \end{subfigure}
     \begin{subfigure}[b]{0.40\linewidth}
    \centering
    \includegraphics[width=\linewidth]{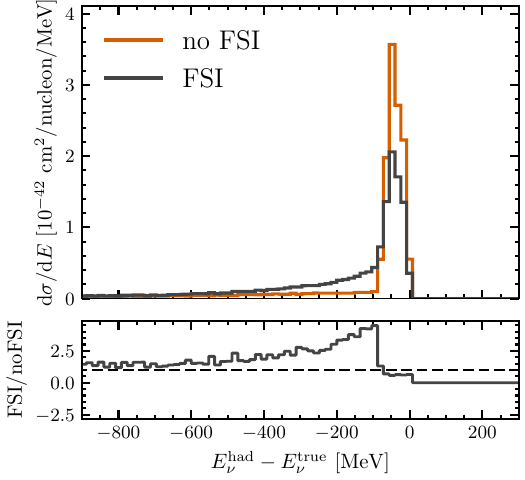}    
    \caption{\footnotesize$E^{\mathrm{had}}_{\nu}$, CCINC-Ar, DUNE $\nu_\mu$ osc. flux}
    \label{fig:EreccalEavailanu}
  \end{subfigure}
  \begin{subfigure}[b]{0.40\linewidth}
    \centering
    \includegraphics[width=\linewidth]{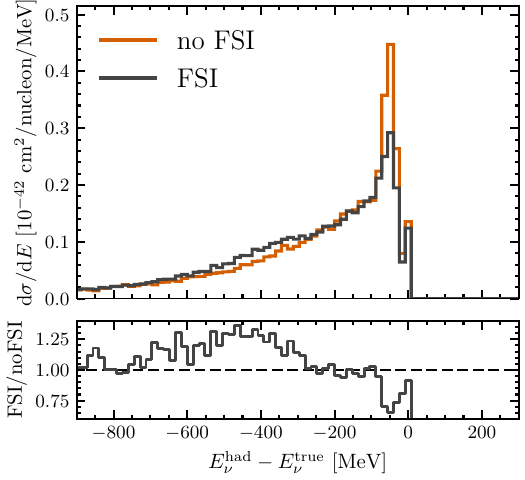}    \caption{\footnotesize$E^{\mathrm{had}}_{\nu}$, CCINC-Ar, DUNE $\bar\nu_\mu$ osc. flux}
    \label{fig:EreccalEhadanu}
  \end{subfigure}
\caption{The NuWro simulated neutrino energy estimation bias (using $E^{\mathrm{had}}_{\nu}$ or $E^{\mathrm{avail}}_{\nu}$) for CC inclusive $\nu_\mu$ or $\bar\nu_\mu$ interactions on an argon target at the DUNE FD using the oscillated DUNE flux (see \autoref{tab:OscProbParams}), split for cases when the NuWro INC is and is not applied.}
\label{fig:Ereccal}
\end{figure}

The neutrino energy biases shown in \autoref{fig:noFSI_HK} and \autoref{fig:noFSI_DUNE} are significantly modified by consideration of FSI as modelled with INCs. This is demonstrated for the Hyper-K and DUNE cases in \autoref{fig:piabs} and \autoref{fig:Ereccal} respectively, where the neutrino energy estimation bias is shown for the NuWro model prediction before and after applying an INC.

In the Hyper-K case, since the neutrino energy estimator is built only from lepton kinematics, that are not affected by the INC by construction, the CCQE and CCnpnh contributions to the neutrino energy bias remain very similar. The most dramatic effect of FSI is instead to add significant extra strength to the distribution where the neutrino energy estimator is $\sim$250-500 MeV smaller than the true neutrino energy. This contribution has similar strength to the total CCnpnh component and stems from pion production interactions in which the outgoing pion is absorbed within the residual nucleus as part of the INC, leaving the final state indistinguishable from CCQE interactions. As with CCnpnh interactions, the bias in the estimator comes from applying \autoref{eq:enuqe} to interactions for which it is not applicable\footnote{In the case of pion-production at Hyper-K, which usually occurs from interactions that excite struck nucleons into resonant states, the assumption that the interaction is with a neutron is therefore incorrect, and the scale of the bias induced is set by the difference between using $m_n$ and the mass of the excited resonance in \autoref{eq:enuqe}.}. There is also a small additional effect from CCQE or CCnpnh interactions where outgoing nucleons stimulate pion production through the cascade, moving interactions out of the signal topology for Hyper-K, but this only affects $\sim1\%$ of these interactions. The impact of the INC is similar for neutrino and antineutrino interactions on complex nuclei (i.e. with $A>1$). However, the antineutrino interactions on hydrogen are not affected by FSI (since free protons are not bound) so the hydrogen peak in \autoref{fig:piabs_anu} remains unchanged from the one in \autoref{fig:noFSIEnuQEanu}. Also, for antineutrinos the slightly smaller portion of CCnpnh events makes the FSI-driven pion absorption contribution to the bias slightly more pronounced. 

\begin{figure}[tbh]
\centering
  \includegraphics[width=0.65\linewidth]{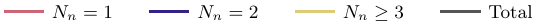}\\[4pt]

  \begin{subfigure}[b]{0.40\linewidth}
    \centering
    \includegraphics[width=\linewidth]{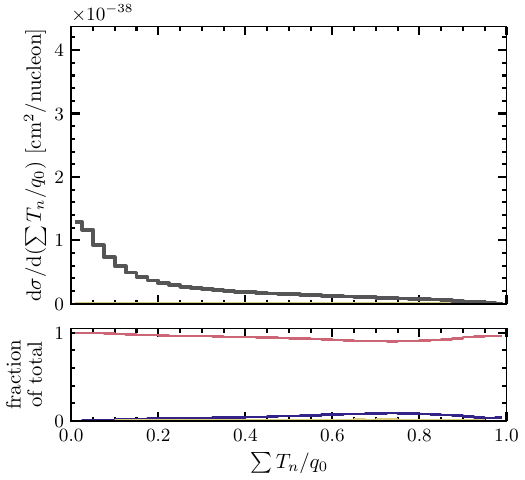}
    \caption{\footnotesize DUNE $\nu_\mu$, no FSI}
  \end{subfigure}
  \begin{subfigure}[b]{0.40\linewidth}
    \centering
    \includegraphics[width=\linewidth]{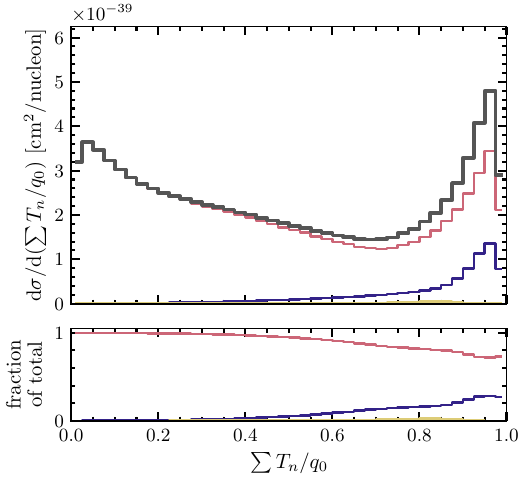}
    \caption{\footnotesize DUNE $\bar\nu_\mu$, no FSI}
  \end{subfigure}\\[6pt]
  \begin{subfigure}[b]{0.40\linewidth}
    \centering
    \includegraphics[width=\linewidth]{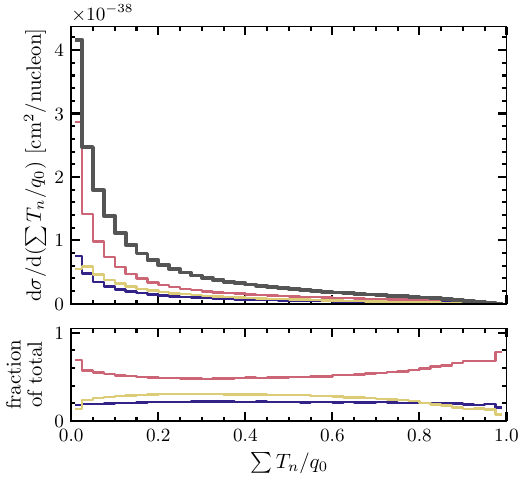}
    \caption{\footnotesize DUNE $\nu_\mu$, with FSI}
  \end{subfigure}
  \begin{subfigure}[b]{0.40\linewidth}
    \centering
    \includegraphics[width=\linewidth]{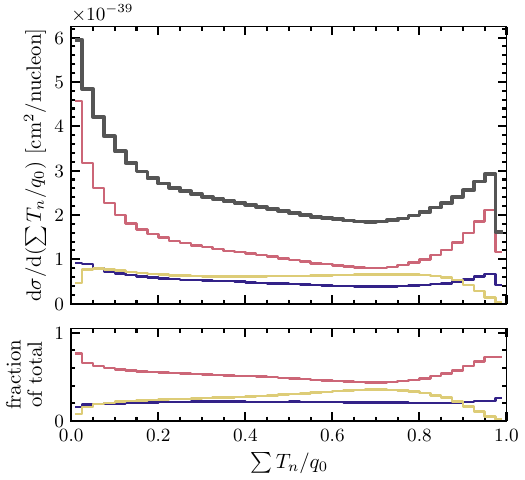}
    \caption{\footnotesize DUNE $\bar\nu_\mu$, with FSI}
  \end{subfigure}
  \caption{The NuWro simulated distribution of the ratio between the sum of final-state neutron kinetic energies ($\Sigma T_n$) and the interaction energy transfer ($q_0$) for CC inclusive $\nu_\mu$ or $\bar\nu_\mu$ interactions on an argon target at the DUNE FD using the oscillated DUNE flux (see \autoref{tab:OscProbParams}). The distribution is split by the number of final state neutrons ($N_n$) in the simulated interactions.}
  \label{fig:neutronsAt_DUNE}
\end{figure}

In the DUNE case, re-scattering of hadrons within the INC alters the outgoing charged pion multiplicity, affecting the bias when using $E^{\mathrm{avail}}_{\nu}$, and the fraction of outgoing energy in neutrons, affecting both $E^{\mathrm{avail}}_{\nu}$ and $E^{\mathrm{had}}_{\nu}$. The impact of the former is most clear in \autoref{fig:EreccalEavailnu}, which shows how the application of the INC changes the relative strength of the discrete peaks corresponding to different charged pion multiplicities. The impact of the INC on the fraction of energy carried away by neutrons is more dramatic. The INC often produces additional neutrons in interactions that otherwise had none, causing a strong migration of interactions away from the low-bias (usually zero-neutron) peak to slightly larger biases. This is further demonstrated in \autoref{fig:neutronsAt_DUNE}, which shows the fraction of energy transferred to the hadronic system in a neutrino interaction ($q_0$) that is given to final-state neutron kinetic energy before and after applying the INC. For neutrino interactions it is clear that FSI simulation via the INC dramatically increases the cross section of interaction topologies containing neutrons, although these neutrons typically take only a small fraction of the energy transfer. This explains the migration of events from the peaks in \autoref{fig:Ereccal} into adjacent higher-bias regions. For antineutrino interactions the cross section of interaction topologies containing neutrons is more modestly increased and the typical fraction of energy transfer carried away by neutrons is actually decreased, explaining the smaller change due to the impact of the INC on the energy estimation bias for antineutrino interactions.

\begin{figure}[tbh]
  \centering
    \includegraphics[width=0.65\linewidth]{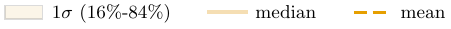}\\[4pt]

    \begin{subfigure}[b]{0.48\linewidth}
      \centering
      \includegraphics[width=\linewidth]{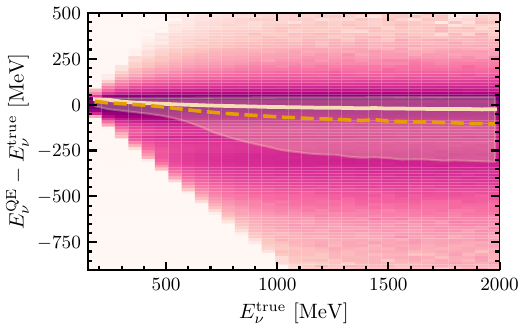}
      \caption{\footnotesize$E^{\mathrm{QE}}_{\nu}$, CC0$\pi$-H$_2$O, Hyper-K $\nu_\mu$ osc.\ flux}
      \label{fig:EdepEnuQEnu}
    \end{subfigure}
    \begin{subfigure}[b]{0.48\linewidth}
      \centering
      \includegraphics[width=\linewidth]{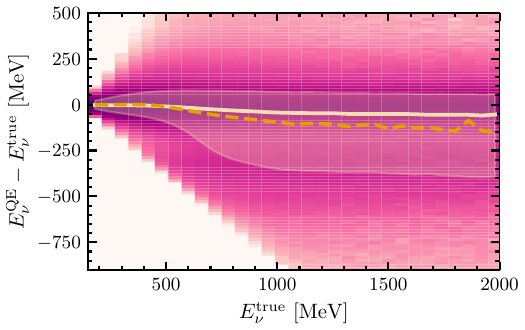}
      \caption{\footnotesize$E^{\mathrm{QE}}_{\nu}$, CC0$\pi$-H$_2$O, Hyper-K $\bar\nu_\mu$ osc.\ flux}
      \label{fig:EdepEnuQEanu}
    \end{subfigure}\\[6pt]

    \includegraphics[width=0.7\linewidth]{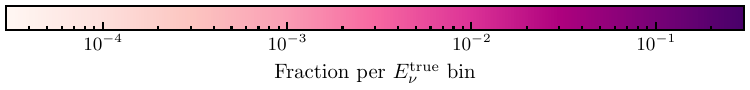}

    \caption{The NuWro simulated neutrino energy estimation bias shape (using $E^{\mathrm{QE}}_{\nu}$) for CC0$\pi$ $\nu_\mu$ or $\bar\nu_\mu$ interactions on a water target as a function of true neutrino energy. The mean, median and a band containing 16-84\% of the bias are shown.}
    \label{fig:FSIEdep_HK}
  \end{figure}

\begin{figure}[tbh]
  \centering
    \includegraphics[width=0.65\linewidth]{Figures/Fig6_legend.pdf}\\[4pt]

    \begin{subfigure}[b]{0.48\linewidth}
      \centering
      \includegraphics[width=\linewidth]{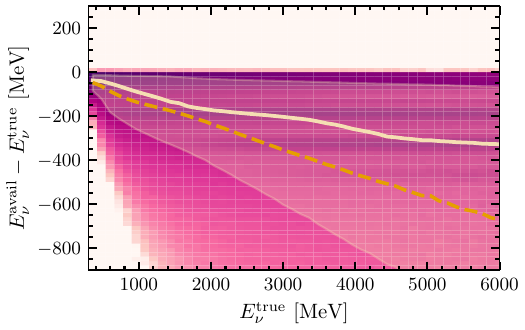}
      \caption{\footnotesize$E^{\mathrm{avail}}_{\nu}$, CCINC-Ar, DUNE $\nu_\mu$ osc.\ flux}
      \label{fig:EdepEnuhadnu}
    \end{subfigure}
    \begin{subfigure}[b]{0.48\linewidth}
      \centering
      \includegraphics[width=\linewidth]{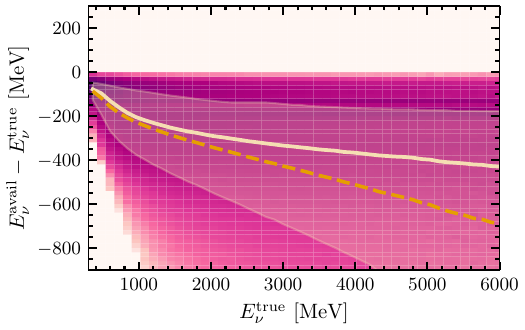}
      \caption{\footnotesize$E^{\mathrm{avail}}_{\nu}$, CCINC-Ar, DUNE $\bar\nu_\mu$ osc.\ flux}
      \label{fig:EdepEnuavailanu}
    \end{subfigure}\\[6pt]

    \begin{subfigure}[b]{0.48\linewidth}
      \centering
      \includegraphics[width=\linewidth]{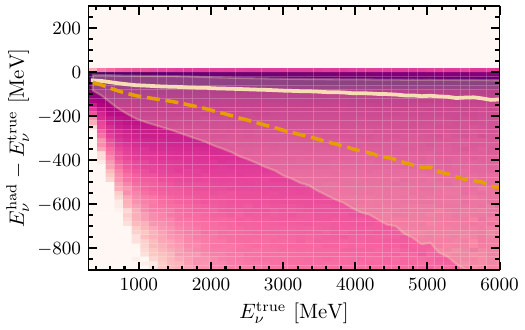}
      \caption{\footnotesize$E^{\mathrm{had}}_{\nu}$, CCINC-Ar, DUNE $\nu_\mu$ osc.\ flux}
      \label{fig:EdepIEnuavailnu}
    \end{subfigure}
    \begin{subfigure}[b]{0.48\linewidth}
      \centering
      \includegraphics[width=\linewidth]{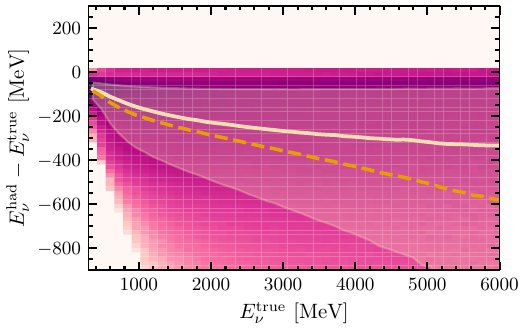}
      \caption{\footnotesize$E^{\mathrm{had}}_{\nu}$, CCINC-Ar, DUNE $\bar\nu_\mu$ osc.\ flux}
      \label{fig:EdepEnuhadanu}
    \end{subfigure}\\[6pt]

    \includegraphics[width=0.7\linewidth]{Figures/Fig6_colorbar_abs_horizontal.pdf}

    \caption{The NuWro simulated neutrino energy estimation bias shape (using $E^{\mathrm{had}}_{\nu}$ or $E^{\mathrm{avail}}_{\nu}$) for CC inclusive $\nu_\mu$ or $\bar\nu_\mu$ interactions on an argon target as a function of true neutrino energy. The mean, median and a band containing 16-84\% of the bias are shown.}
  \label{fig:FSIEdep_DUNE}
\end{figure}

\autoref{fig:piabs} and \autoref{fig:Ereccal} show neutrino energy estimation bias averaged over the oscillated muon (anti)neutrino flux for Hyper-K and DUNE. However, the size and shape of the bias varies with neutrino energy itself and so may differ for variations of neutrino oscillation parameters and between the ND and FD. To illustrate the magnitude of this effect, \autoref{fig:FSIEdep_HK} and \autoref{fig:FSIEdep_DUNE} show the evolution of the neutrino energy estimation bias as a function of neutrino energy for the Hyper-K and DUNE cases respectively. It is clear that larger biases are more likely when neutrino energies are larger, primarily driven by the changing contribution from different interaction channels (as demonstrated in \autoref{fig:sigmaEnu}). For Hyper-K this is because the bias stems from CCnpnh and pion-producing interactions which both have cross sections that grow faster than the CCQE one above $\sim$0.5~GeV neutrino energy, making their relative contribution to CC0$\pi$ interactions increase above this. For DUNE a similar argument applies where higher energy transfer interactions, which can produce higher energy neutrons and higher charged pion multiplicities, become more likely at larger neutrino energies. The same qualitative conclusions apply for both neutrinos and antineutrinos. 

\subsection{FSI beyond the INC}
\label{subsec:beyondcasc}

\begin{figure}[tbh]
 \centering
   \begin{subfigure}[b]{0.32\linewidth}
    \centering
    \includegraphics[width=\linewidth]{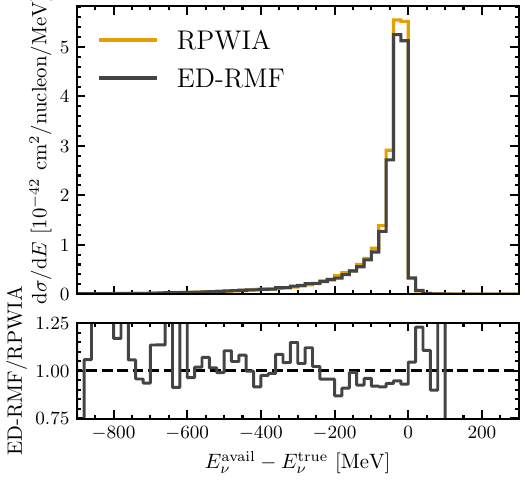}    \caption{\tiny$E^{\mathrm{avail}}_{\nu}$, CCQE-Ar, DUNE $\nu_\mu$ osc. flux}
    \label{fig:EDRMFEnuhadnu}
  \end{subfigure}
  \begin{subfigure}[b]{0.32\linewidth}
    \centering
    \includegraphics[width=\linewidth]{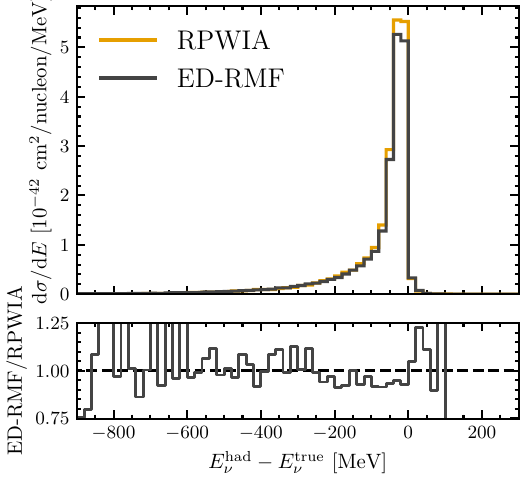}
    \caption{\tiny$E^{\mathrm{had}}_{\nu}$, CCQE-Ar, DUNE $\nu_\mu$ osc. flux}
    \label{fig:EDRMFEnuavailnu}
  \end{subfigure}
  \begin{subfigure}[b]{0.32\linewidth}
    \centering
    \includegraphics[width=\linewidth]{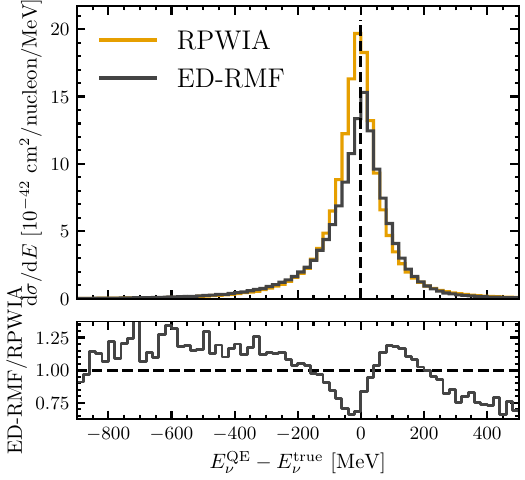}        \caption{\tiny$E^{\mathrm{QE}}_{\nu}$, CCQE-H$_2$O, Hyper-K $\nu_\mu$ osc. flux}
    \label{fig:EDRMFEnuQEnu}
  \end{subfigure}

 \caption{The NEUT simulated neutrino energy estimation bias (using $E^{\mathrm{had}}_{\nu}$, $E^{\mathrm{avail}}_{\nu}$ or $E^{\mathrm{QE}}_{\nu}$) for CCQE $\nu_\mu$ interactions on an argon or water target at the DUNE or Hyper-K FD using the oscillated DUNE or Hyper-K flux (see \autoref{tab:OscProbParams}) and the ED-RMF or RPWIA models.}
\label{fig:rmfcomp}
\end{figure}

As discussed in \autoref{sec:methods}, the semi-classical approach of INCs does not consider the full impact of FSI that is captured by modern microscopic models. The latter consider an additional facet of FSI through the incorporation of a nuclear potential that distorts the outgoing nucleon wavefunction in the cross-section calculation, as described in \autoref{subsec:edrmf_model}. The impact of this potential can be quantified by comparing the neutrino energy estimator bias for NEUT simulations using the ED-RMF model (which includes this potential) and the RPWIA model (which neglects it but is otherwise equivalent). This is shown for neutrino CCQE interactions in \autoref{fig:rmfcomp}, for an argon and water target for the cases of DUNE and Hyper-K respectively\footnote{For neutrino interactions, no CCQE interactions can occur on hydrogen atoms, so all interactions shown here are in fact on oxygen.}. The RPWIA model using an argon target includes only neutrino inputs at the time of writing; therefore, only neutrino interactions are generated here. 

Since the consideration of the nuclear potential in ED-RMF does not directly stimulate neutron or pion production, the DUNE estimators are not very sensitive to differences between ED-RMF and RPWIA. Conversely, \autoref{eq:enuqe} is very sensitive to the alteration of interaction kinematics for CCQE interactions between ED-RMF and RPWIA, causing a significant shift in the bias distribution. This shift originates from the nuclear mean-field potential, which modifies the initial- and final-state nucleon kinematics, indirectly altering the scattered lepton kinematics through energy-momentum conservation. In the RPWIA case, this does not occur and \autoref{eq:enuqe} is expected to be symmetric about zero bias. Similar effects have previously been reported in Ref.~\cite{Ankowski:2014yfa}.

\FloatBarrier

\subsection{FSI model variations}
\label{subsec:fsivar}

\autoref{fig:piabs} and \autoref{fig:Ereccal} demonstrate large differences in neutrino energy estimation bias between cases where FSI (or some aspect of it) is or is not modelled. This is an overly conservative modification, as the need for FSI is well established. Here we consider a number of more realistic variations of FSI modelling, and quantify the impact each has on neutrino energy estimation bias, covering:

\begin{itemize}
    \item Variations of the pion absorption probability within the NuWro INC. We performed a minimal change to the NuWro source code by introducing a single parameter which directly scales the pion absorption cross section used in the INC. Prior attempts to tune INC models to pion-scattering measurements~\cite{PinzonGuerra:2018rju} suggest significant uncertainty on the probability for pion absorption. Based on the prescription in Refs.\cite{PinzonGuerra:2018rju, T2K:2023smv} we vary the pion absorption interaction cross section within the NuWro INC by $\pm$31\%. A change of a similar size is prescribed by findings in Ref.~\cite{Dytman:2021ohr}.
    \item Variations of the nucleon interaction probability inside the NuWro INC. Based on uncertainties in measurements of nucleon transparency from electron scattering~\cite{CLAS:2025fqh,Niewczas:2019fro}, we consider variations of the nominal NuWro simulation in which the mean free path (MFP) between interactions of nucleons within the INC is varied by $\pm$30\% using existing NuWro parameters. 
    \item Variations of the entire INC model used. Different INCs make different assumptions regarding propagation of hadrons, the kinematics of scatters and the role of nuclear effects, and tune the relative importance of processes using different data sets. We vary the INC considered among the four models available within GENIE discussed in \autoref{sec:methods}. The impact of the variation of these models on DUNE neutrino energy reconstruction is also discussed in Ref.~\cite{Liu:2025hpl}.
\end{itemize}

In the cases where we vary the pion absorption probability and the nucleon mean free path, the size of the variation is extracted from measurements on light nuclei such as carbon and oxygen. For the purpose of this work we apply the same variation for interactions on an argon target. Future analyses, for example exploiting recent measurements of pion-argon and proton-argon scattering such as Ref.~\cite{LArIAT:2021yix,DUNE:2025zhx,DUNE:2025pda}, may provide different ranges for varying these parameters in future studies.

\begin{figure}[tbp]
\centering
  \includegraphics[width=\linewidth]{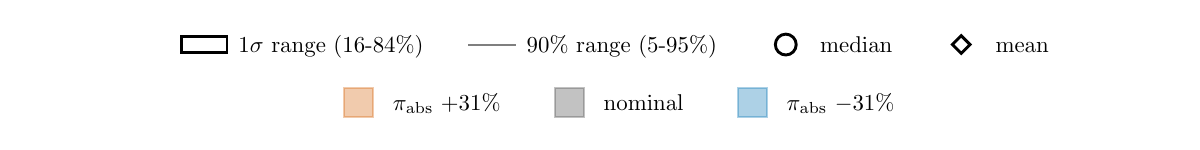}\\[4pt]
  \begin{subfigure}[b]{0.49\linewidth}
    \centering
    \includegraphics[width=\linewidth]{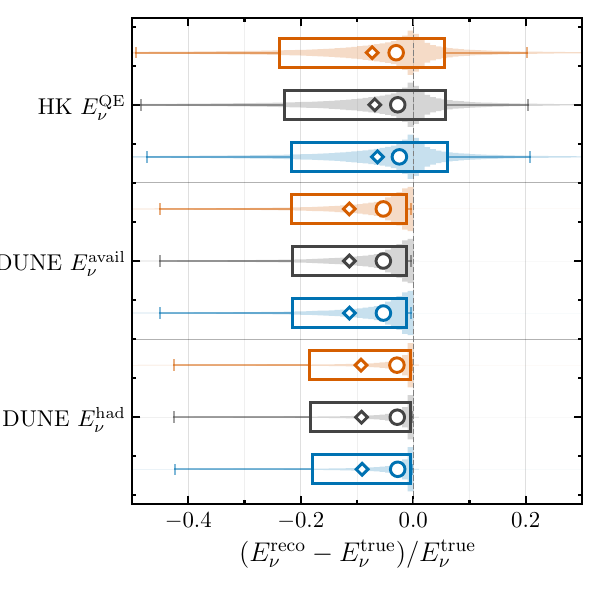}
    \caption{\footnotesize $\nu_\mu$}
  \end{subfigure}%
  \begin{subfigure}[b]{0.49\linewidth}
    \centering
    \includegraphics[width=\linewidth]{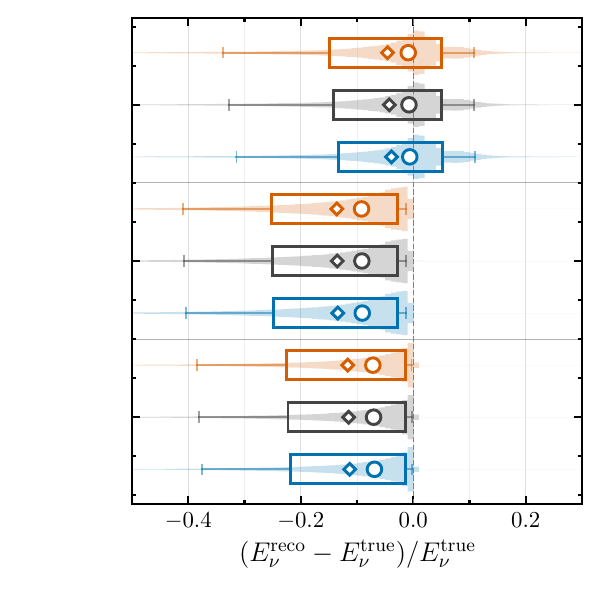}
    \caption{\footnotesize $\bar\nu_\mu$}
  \end{subfigure}
  \caption{The relative neutrino energy estimation bias, $(E^{\mathrm{reco}}_{\nu} - E^{\mathrm{true}}_{\nu})/E^{\mathrm{true}}_{\nu}$, for the Hyper-K case (where  $E^{\mathrm{reco}}_{\nu}=E^{\mathrm{QE}}_{\nu}$), and DUNE cases (where  $E^{\mathrm{reco}}_{\nu}=E^{\mathrm{avail}}_{\nu}$ or $E^{\mathrm{reco}}_{\nu}=E^{\mathrm{had}}_{\nu}$), for NuWro simulations with the pion absorption probability modified by $\pm31\%$. The silhouette shows the shape of the energy estimation bias, overlaid with the median (open circle) and mean (open diamond) bias, as well as a box giving the 1$\sigma$ (16-84\%) range of the distribution and faint 90\% whiskers (5-95\%).}
  \label{fig:bw_piabs}
\end{figure}

\begin{figure}[tbp]
\centering
  \includegraphics[width=\linewidth]{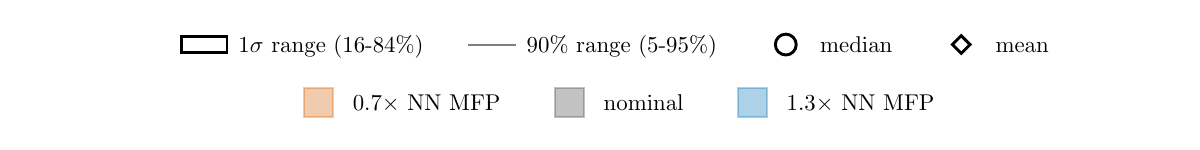}\\[4pt]
  \begin{subfigure}[b]{0.49\linewidth}
    \centering
    \includegraphics[width=\linewidth]{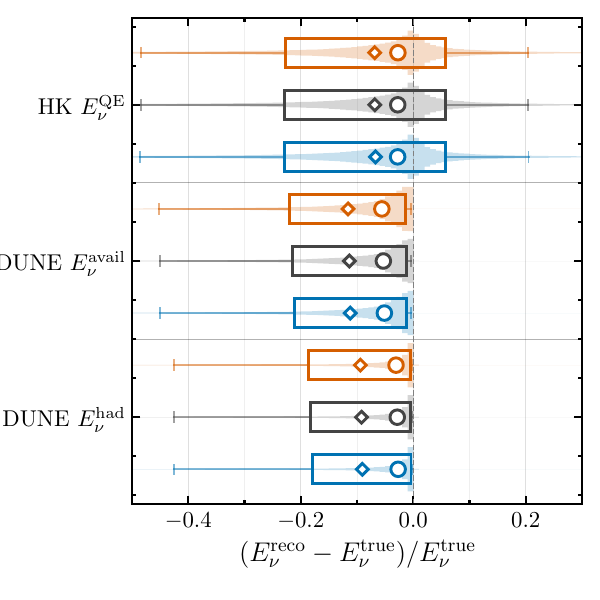}
    \caption{\footnotesize $\nu_\mu$}
  \end{subfigure}%
  \begin{subfigure}[b]{0.49\linewidth}
    \centering
    \includegraphics[width=\linewidth]{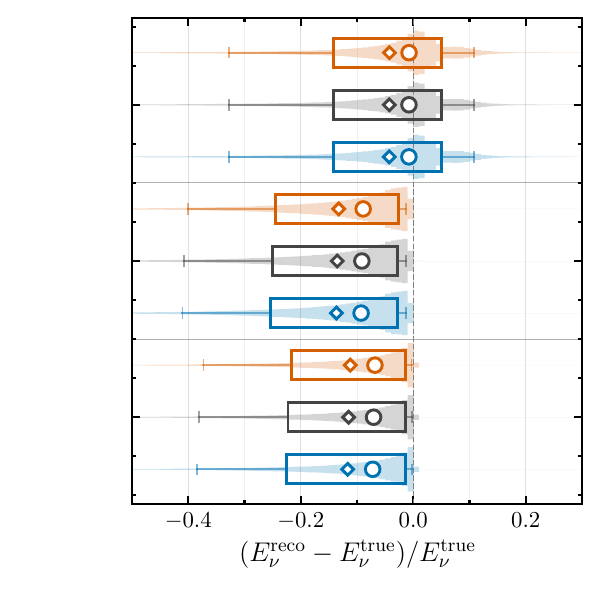}
    \caption{\footnotesize $\bar\nu_\mu$}
  \end{subfigure}
  \caption{The same as Fig.~\ref{fig:bw_piabs} but considering $\pm$30\% alterations to the nucleon MFP within the INC.}
  \label{fig:bw_mfp}
\end{figure}

\begin{figure}[htb]
\centering
  \includegraphics[width=\linewidth]{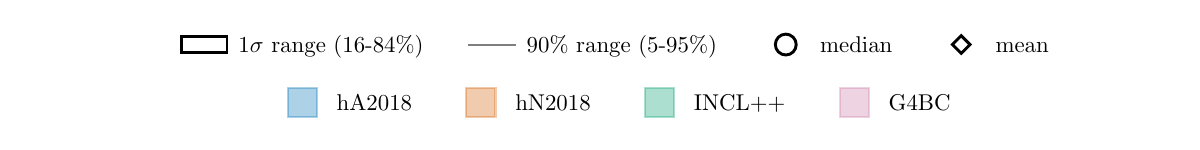}\\[4pt]
  \begin{subfigure}[b]{0.49\linewidth}
    \centering
    \includegraphics[width=\linewidth]{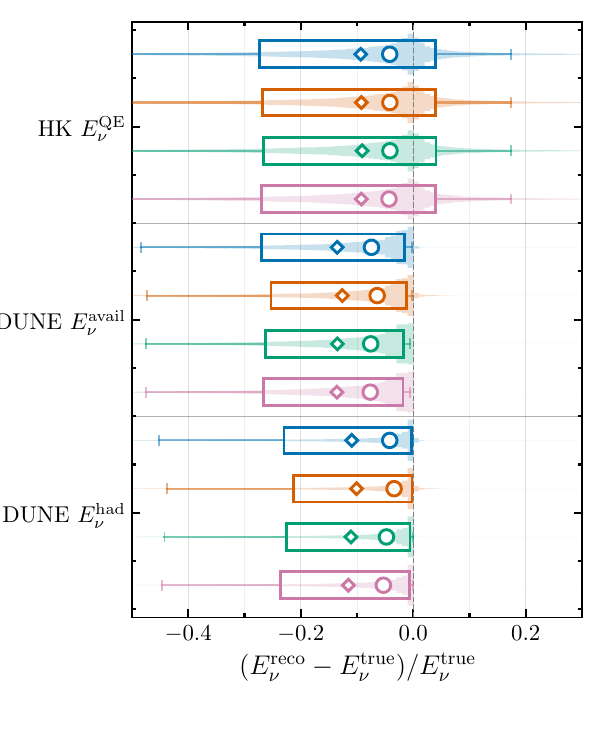}
    \caption{\footnotesize $\nu_\mu$}
  \end{subfigure}%
  \begin{subfigure}[b]{0.49\linewidth}
    \centering
    \includegraphics[width=\linewidth]{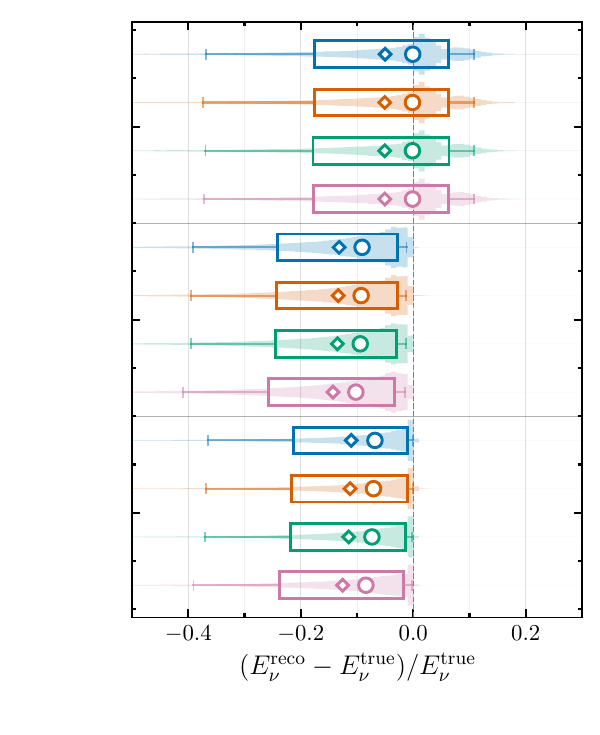}
    \caption{\footnotesize $\bar\nu_\mu$}
  \end{subfigure}
  \caption{The same as Fig.~\ref{fig:bw_piabs} but considering four GENIE INC model variations.}
  \label{fig:bw_genie}
\end{figure}

\autoref{fig:bw_piabs}, \autoref{fig:bw_mfp}, and \autoref{fig:bw_genie} show the difference between the true neutrino energy and each neutrino energy estimator for each of the three FSI model variations. For easier comparison between the Hyper-K and DUNE cases, the neutrino energy estimation bias \textit{relative} to the true neutrino energy is shown. Similar figures considering the impact of simply turning FSI on and off are shown in \autoref{app:moreFSIvar}. For the DUNE and Hyper-K neutrino and antineutrino cases, the figures illustrate how the mean, median and spread of the neutrino energy estimation bias changes for each variation.

The change in the mean and median estimation bias for each variation is summarised in \autoref{tab:FSIVar}, and the same shifts expressed relative to the true neutrino energy are provided in \autoref{app:relBias}. Whilst variations of the mean are most indicative of an energy estimation scale shift comparable with the $\sim$5 MeV and $\sim$15 MeV requirements for ultimate precision in \autoref{sec:enurec}, a comparison of the mean and median helps to show how the energy estimation bias is reshaped by FSI as well as shifted on average. The table also includes the impact of turning FSI on and off as well as the impact of applying a nuclear potential (as described in \autoref{subsec:beyondcasc}). For the latter, note that only $\nu_\mu$ CCQE interactions are simulated. 

Since pion absorption is the dominant driver of neutrino energy estimation bias from INC FSI for Hyper-K, varying its strength almost directly varies the size of the FSI/no-FSI differences shown in \autoref{fig:piabs}. For DUNE the impact is less pronounced. Whilst the absorption of pions moves events between the discrete pion peaks shown in \autoref{fig:EreccalEavailnu}, it also redistributes some of the absorbed pion kinetic energy to outgoing neutrons.

Varying the nucleon MFP has very little impact on Hyper-K neutrino energy estimation, as outgoing nucleons are not measured. The only effect is to marginally alter the small probability for pion production within the INC, migrating some CCQE interactions out of the CC0$\pi$ sample of interactions considered. For DUNE, a smaller (larger) MFP nucleon increases (decreases) the probability for protons to transfer energy to neutrons. For neutrino interactions, which produce more protons than neutrons at the interaction vertex, the impact of a lower MFP is therefore to redistribute energy from protons to neutrons, increasing the neutrino energy estimation bias. For antineutrino interactions, which contain more neutrons than protons, the effect is the opposite. 

Alterations to the INC model itself implicitly change both the pion absorption probability and the nucleon MFP at the same time as varying the kinematics of each process in the INC (e.g. the amount of momentum transferred to protons or neutrons in a typical pion absorption). Since Hyper-K is predominantly sensitive to only the probability of pion absorption (and not the kinematics of the process), and the INCs all have similar pion absorption probabilities (likely from tuning to similar pion scattering data sets), the impact on Hyper-K neutrino energy estimation is smaller than for DUNE. It is not negligible, however: the median bias shifts by less than 1~MeV, but the mean shifts by up to $\sim$9~MeV. Because each pion-absorption event carries a large bias ($\sim$250--500~MeV, \autoref{fig:piabs}), even small differences in pion absorption probability between the INC models move the mean of the bias distribution while leaving its median almost unchanged. For DUNE, altering the INC can change the fraction of energy carried away by neutrons or the pion multiplicity in complicated ways, leading to a significant impact on the neutrino energy estimation bias. The INC variations affect the DUNE-case neutrino and antineutrino energy estimators similarly.

\begin{table}[htb]
\centering
\scriptsize
\setlength{\tabcolsep}{4pt}
\renewcommand{\arraystretch}{1.15}
\begin{tabular}{@{}c l r r r r r r r r r r@{}}
\toprule
\textbf{Flavour} & \textbf{Observable} & \multicolumn{2}{c}{\textbf{FSI / no FSI}} & \multicolumn{2}{c}{\textbf{$\pi_{\mathrm{abs}}\pm31\%$}} & \multicolumn{2}{c}{\textbf{NN MFP $\pm30\%$}} & \multicolumn{2}{c}{\textbf{INC model var.}} & \multicolumn{2}{c}{\textbf{Nuc. Pot. on/off}} \\
\cmidrule(lr){3-4}\cmidrule(lr){5-6}\cmidrule(lr){7-8}\cmidrule(lr){9-10}\cmidrule(lr){11-12}
 &  & median & mean & median & mean & median & mean & median & mean & median & mean \\
 &  & [MeV] & [MeV] & [MeV] & [MeV] & [MeV] & [MeV] & [MeV] & [MeV] & [MeV] & [MeV] \\
\midrule
\multirow{3}{*}{\rotatebox[origin=c]{90}{\textbf{$\nu_\mu$}}} & Hyper-K $E_{\nu}^{\mathrm{QE}}$ & \cellcolor{red!12}10.1 & \cellcolor{red!12}35.8 & \cellcolor{red!12}5.8 & \cellcolor{red!12}16.4 & 0.2 & 1.0 & 0.8 & \cellcolor{red!12}9.4 & \cellcolor{red!12}7.0 & 4.4 \\
 & DUNE $E_{\nu}^{\mathrm{avail}}$ & \cellcolor{red!12}28.7 & \cellcolor{red!12}79.4 & 1.6 & 2.9 & 10.6 & 4.2 & \cellcolor{red!12}44.5 & \cellcolor{red!12}32.2 & 0.3 & 4.5 \\
 & DUNE $E_{\nu}^{\mathrm{had}}$ & \cellcolor{red!12}88.7 & \cellcolor{red!12}81.9 & 13.8 & 10.7 & \cellcolor{red!12}16.3 & 2.4 & \cellcolor{red!12}39.2 & \cellcolor{red!12}27.1 & 0.3 & 4.1 \\
\midrule
\multirow{3}{*}{\rotatebox[origin=c]{90}{\textbf{$\bar\nu_\mu$}}} & Hyper-K $E_{\nu}^{\mathrm{QE}}$ & \cellcolor{red!12}7.0 & \cellcolor{red!12}25.6 & 3.6 & \cellcolor{red!12}12.5 & 0.3 & 1.1 & 0.1 & 2.0 &  &  \\
 & DUNE $E_{\nu}^{\mathrm{avail}}$ & \cellcolor{red!12}39.8 & \cellcolor{red!12}35.4 & 7.7 & 6.5 & 14.7 & \cellcolor{red!12}16.2 & \cellcolor{red!12}39.8 & \cellcolor{red!12}16.4 &  &  \\
 & DUNE $E_{\nu}^{\mathrm{had}}$ & \cellcolor{red!12}55.1 & \cellcolor{red!12}43.0 & \cellcolor{red!12}17.8 & 13.7 & \cellcolor{red!12}17.3 & \cellcolor{red!12}18.7 & \cellcolor{red!12}61.7 & \cellcolor{red!12}37.0 &  &  \\
\bottomrule
\end{tabular}
\normalsize
\caption{The maximum shift in the mean and median neutrino energy estimation bias due to different FSI variations for the Hyper-K and DUNE neutrino and antineutrino cases. The numbers reported for the INC model variation is derived from the two INC models that give the largest spread in the mean. Red boxes indicate that the variation is larger than 5 MeV or 15 MeV for the Hyper-K and DUNE cases respectively, which is broadly indicative of how well the neutrino energy reconstruction scale must be controlled (see \autoref{sec:enurec}). ``Nuc. Pot.'' stands for the nuclear potential considered in \autoref{subsec:beyondcasc}, for which the table reports a shift derived considering only CCQE interactions.}
\label{tab:FSIVar}
\end{table}

\FloatBarrier

\FloatBarrier

\section{Discussion}
\label{sec:discussion}

The analysis presented in \autoref{sec:results} clearly demonstrates that FSI has the potential to be an important driver of neutrino energy estimation bias for both DUNE and Hyper-K. Several of the realistic variations of FSI models cause shifts in the median and mean estimation bias significantly exceeding the $\sim$5~MeV and $\sim$15~MeV precision requirements. Beyond this overall conclusion, the results also reveal that FSI affects the two experiments in fundamentally different ways that follow directly from the differences in the experiments' energy estimators. 

The Hyper-K $E^{\mathrm{QE}}_{\nu}$ estimator is built from lepton kinematics alone and is therefore blind to any INC effect that does not change the probability for a neutrino interaction to leave a CC0$\pi$ topology. As discussed in \autoref{subsec:impactofINC}, the only INC effect that significantly biases $E^{\mathrm{QE}}_{\nu}$ is pion absorption, which moves resonant pion production events, for which \autoref{eq:enuqe} produces a bias, into the CC0$\pi$ topology. Conversely, INC-driven changes to nucleon kinematics or charged pion multiplicities above zero are invisible to the Hyper-K energy estimator. The DUNE $E^{\mathrm{had}}_{\nu}$ estimator behaves in the opposite way: it is largely insensitive to topology migrations of the CC0$\pi$/CC1$\pi$ type but is strongly affected by any aspect of FSI that alters the fraction of the hadronic energy carried by neutrons. This leaves the estimator particularly sensitive to variations of the different GENIE INC models, whilst showing more modest sensitivity to variations of the nucleon MFP and pion absorption (the latter being more impacted by energy transferred to neutrons during the absorption process than because of the absorption itself). The DUNE $E^{\mathrm{avail}}_{\nu}$ estimator is additionally sensitive to the multiplicity of charged pions.  Interestingly, at least for the NuWro INC, this actually reduces its sensitivity to changes in pion absorption. This is because the absorption of a pion reduces bias which actually counteracts the increase in bias from any potential kinetic energy transferred to neutrons during the absorption. 

The results in \autoref{subsec:beyondcasc} demonstrate that FSI effects beyond those considered in an INC are important, especially for Hyper-K. Whilst incorporating the nuclear mean-field potential causes a small shift in the mean of the $E^{\mathrm{QE}}_{\nu}$ bias distribution for CCQE interactions, the distribution is significantly reshaped, as shown by the $\sim$7~MeV median shift and \autoref{fig:rmfcomp}. For DUNE, the role of the nuclear potential in CCQE interactions seems less pronounced. However, an analogous microscopic treatment for pion production could in principle have a more direct impact on calorimetric estimators, and quantifying this remains an open challenge.

A further feature of the variations summarised in \autoref{tab:FSIVar} is that FSI modelling  does not generally affect neutrino and antineutrino energy estimation symmetrically. In particular, the nucleon MFP variations for DUNE cause shifts in the opposite direction for neutrinos and antineutrinos. Moreover, shifts from INC model variations can reshape the energy estimation bias for the DUNE estimators quite differently for neutrinos and antineutrinos, as seen by the different impact on the bias medians. The implication for oscillation analyses may be significant, since sensitivity of $\delta_{\mathrm{CP}}$ is at least partially extracted from differences between the oscillated neutrino and antineutrino spectra, which may thus be confused by FSI mismodelling. 

Importantly, \autoref{fig:FSIEdep_HK} and \autoref{fig:FSIEdep_DUNE}  show that neutrino energy estimation biases introduced by FSI are not constant in neutrino energy. The figure demonstrates that both kinematic and calorimetric biases grow with $E_\nu$, driven for Hyper-K by the rising contribution of CCnpnh and pion production relative to CCQE above $\sim$0.5~GeV, and for DUNE by the increasing typical energy transfer and the associated growth in neutron emission and charged-pion multiplicity. This energy dependence has an important implication for the use of ND measurements to constrain the size of the bias. The ND and FD see different neutrino energy spectra (since the latter is shaped by oscillations), the FSI-driven bias does not compensate between the two detectors, and an ND constraint on the bias necessarily relies on the underlying interaction model to extrapolate across energies. However, this may be partially mitigated through use of movable ND and the PRISM technique~\cite{nuPRISM:2014mzw, DUNE:2021tad, Hyper-Kamiokande:2025asb}.

\begin{figure}[tb]
\centering
  \begin{subfigure}[b]{0.4\linewidth}
    \centering
    \includegraphics[width=\linewidth]{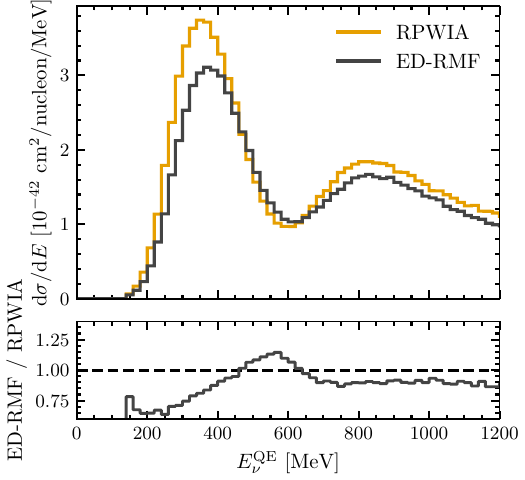}
    \caption{\footnotesize$E^{\mathrm{true}}_{\nu}$, CCQE-H$_2$O, Hyper-K $\nu_\mu$ osc.\ flux}
    \label{fig:edrmf_osc_numu}
  \end{subfigure}%
  \begin{subfigure}[b]{0.4\linewidth}
    \centering
    \includegraphics[width=\linewidth]{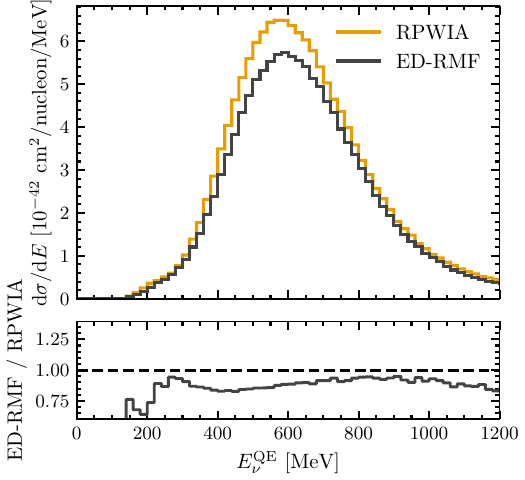}
    \caption{\footnotesize$E^{\mathrm{true}}_{\nu}$, CCQE-H$_2$O, Hyper-K $\nu_e$ osc.\ flux}
    \label{fig:edrmf_osc_nue}
  \end{subfigure}
  \caption{The NEUT simulated rate of CCQE $\nu_\mu$ or $\nu_e$ interactions on water using the oscillated Hyper-K flux (see \autoref{tab:OscProbParams}) as a function of estimated neutrino energy ($E_{\nu}^\mathrm{QE}$) using the ED-RMF and RPWIA models.}
  \label{fig:edrmf_osc}
\end{figure}

The richness of the FSI-related effects on neutrino energy estimation is difficult to capture with simple summary metrics, such as the mean and median reported in \autoref{tab:FSIVar}. \autoref{fig:edrmf_osc} illustrates this for the change of nuclear potential in NEUT (the difference between ED-RMF and RPWIA). Although this variation yields a negligible shift in the mean according to \autoref{tab:FSIVar}, it produces a much larger median shift and substantially reshapes the CCQE component of the spectrum. This disparity between the mean and median shifts is a direct consequence of the asymmetric way the nuclear potential reshapes the distribution.

Throughout this work, we have primarily considered the bias in the neutrino energy estimators averaged over the oscillated $\nu_\mu$ or $\bar\nu_\mu$ energy spectra. However, the appearance spectra for $\nu_e$ and $\bar\nu_e$ have a significantly different shape than the $\nu_\mu$ and $\bar\nu_\mu$ spectra, as shown in \autoref{fig:sigmaEnu} and discussed in \autoref{subsec:osc_spectra}. Since the bias itself varies with neutrino energy, as demonstrated in \autoref{fig:FSIEdep_HK} and \autoref{fig:FSIEdep_DUNE}, this different spectral shape leads to a different flux-averaged result. In addition, although the way that $\nu_\mu$ and $\nu_e$ cross sections differ is well known and (assuming lepton universality) driven only by the lepton mass, the flux-averaged $\nu_e$ and $\nu_\mu$ event rates remain susceptible to uncertainties in how poorly understood nuclear effects convolve with the available interaction phase space, which differs between the two species~\cite{Dieminger:2023oin, Ankowski:2017yvm, Nikolakopoulos:2019qcr, Martini:2023kem}. We have nevertheless checked that the observations made for $\nu_\mu$ and $\bar\nu_\mu$ interactions remain similar in the $\nu_e$ and $\bar\nu_e$ case. The most notable difference is that the DUNE INC model spread increases slightly for both neutrinos and antineutrinos. The scale of the mean and median variations due to FSI modelling remains similar to that presented in \autoref{tab:FSIVar} and the discussion based on the $\nu_\mu$ and $\bar\nu_\mu$ flux-averaged results is representative for the case of $\nu_e$ and $\bar\nu_e$ appearance.

Overall, the results of \autoref{sec:results} clearly demonstrate the importance of FSI modelling for DUNE and Hyper-K, but the shifts presented in \autoref{tab:FSIVar} should not be read as quantitatively indicative of systematic uncertainties that the experiments will face. Firstly, the variations shown are raw model spreads, presented without any constraint from ND measurements. In a real oscillation analysis, FSI parameters would be treated as nuisance parameters and constrained through fits to ND data, so that the residual bias propagated to the FD would be substantially smaller than the raw spread. The power of ND constraints will depend on detector performance and the development of a neutrino interaction model, and associated uncertainties, that can describe the ND data. Secondly, we have considered the impact of FSI on neutrino energy estimation in isolation from other observables. In practice, FD measurements can be binned in additional kinematic variables beyond estimated neutrino energy (for example visible hadronic energy, or observed proton and pion multiplicities), which can provide additional handles to disentangle FSI-driven migrations from genuine oscillation effects. The extent to which such multi-dimensional analyses can mitigate the biases shown here is a question we do not address in this work. Thirdly, how these energy-estimation shifts ultimately translate into biases on the measured oscillation parameters is impossible to assess quantitatively without a full oscillation analysis. As \autoref{fig:edrmf_osc} shows, even a variation with a negligible mean shift can move the oscillation dip (by $\sim$18~MeV for the nuclear-potential change) and alter the normalisation of the oscillated spectra (unlike the INC variations, which are designed to conserve the inclusive cross section as a function of lepton kinematics). In summary, \autoref{tab:FSIVar} should be read as characterising the size of the FSI modelling problem to be solved and not the size of the residual bias entering a final oscillation analysis. 

There are some clear opportunities to confront the challenges in FSI modelling faced by next generation neutrino oscillation experiments. For Hyper-K, a key priority is to ensure that pion absorption on oxygen is well modelled and is accompanied by robust uncertainties that can be constrained at the ND. This would be greatly facilitated by new pion-oxygen scattering measurements to supplement the limited available data at relevant kinematics for Hyper-K~\cite{PinzonGuerra:2018rju}, which could be used to directly benchmark INC models. In parallel, microscopic FSI treatments must also be accompanied by uncertainties to allow an ND constraint and ideally extended beyond CCQE so that the ED-RMF-type shifts seen in this work can be assessed for the CCnpnh and pion-production channels. For DUNE, the priority is to constrain the repartition of hadronic energy between neutrons, protons and pions caused by FSI. New proton- and pion-argon scattering measurements from ProtoDUNE, building on its existing results~\cite{DUNE:2025pda, DUNE:2025zhx}, may provide an important constraint, especially if the energy transferred to neutrons can be inferred through exclusive, differential analyses. Moreover, with high precision detectors and large projected statistics, measurements from the DUNE ND~\cite{DUNE:2021tad} itself will be well placed to constrain FSI modelling, provided sufficient work has been done to develop a parameterised neutrino interaction model that can be constrained. In both the DUNE and Hyper-K cases, the path to controlling FSI at the level to allow the experiments to reach their ultimate sensitivity requires coupling theoretical developments to a dedicated programme of experimental measurements.

\section{Conclusions}
\label{sec:conclusions}

In this work, we have used state-of-the-art neutrino interaction event generators to characterise the impact of FSI modelling on the neutrino energy estimators that DUNE and Hyper-K will use to extract oscillation parameters, considering both the semi-classical INC that dominate current simulations and a microscopic treatment based on a relativistic nuclear potential. We have shown that the target precisions on oscillation parameter measurements quoted by Hyper-K and DUNE correspond broadly to a control of the estimated neutrino energy scale at the level of $\sim$5~MeV and $\sim$15~MeV respectively. Realistic variations of FSI models produce shifts in neutrino energy estimation bias that are at or above these target energy scale precisions. The two experiments are sensitive to different FSI physics: the kinematic estimator used at Hyper-K is driven primarily by the probability of pion absorption and by the consideration of a mean-field nuclear potential outside the semi-classical INC paradigm; the calorimetric estimators used at DUNE are driven by the distribution of hadronic energy between neutrons, protons and pions. 

The variations reported here should be understood as characterising the size of the FSI modelling problem to be solved, rather than as residual systematic uncertainties on an oscillation analysis, which will benefit from ND constraints. The path to controlling FSI at the level demanded by DUNE and Hyper-K requires coordinated progress on two fronts: continued theoretical development of FSI models, including the extension of microscopic treatments to channels beyond CCQE and the construction of robust parameterised uncertainties that can be propagated through oscillation analyses; and a dedicated programme of experimental measurements targeted at the specific aspects of FSI to which each experiment is most sensitive. 

\ack{The authors thank R. Dharmapal Banerjee for help with the NuWro event generator, C. Wilkinson for having produced the GENIE simulation files used for this work and L. Pickering for providing software containers including NUISANCE and neutrino event generators used in early iterations of this work. J.~McKean was supported by Grant-in-Aid for JSPS Research Fellows, JSPS KAKENHI Grant No. 25KF0223.}

\appendix

\section{Impact of the neutrino energy estimation bias on antineutrino oscillation spectra}
\label{app:anu}

In this appendix, we perform an analogous study to the one presented in \autoref{subsec:osc_spectra}, but where we focus on the impact of oscillations and energy scale changes on \textit{antineutrino} FD spectra.

\begin{figure}[htb]
\centering
  \includegraphics[width=\linewidth]{Figures/legend_Fig1_HK_dm32.pdf}\\[2pt]
  \begin{subfigure}[b]{0.40\linewidth}
    \centering
    \includegraphics[width=\linewidth]{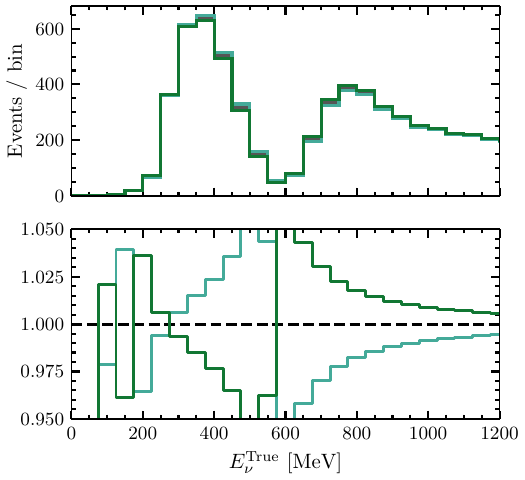}
    \caption{\footnotesize$E^{\mathrm{true}}_{\nu}$, CC0$\pi$-H$_2$O, Hyper-K $\bar{\nu}_\mu$ osc.\ flux}
    \label{fig:Enuspect_anu_dm32_HK_true}
  \end{subfigure}
  \begin{subfigure}[b]{0.40\linewidth}
    \centering
    \includegraphics[width=\linewidth]{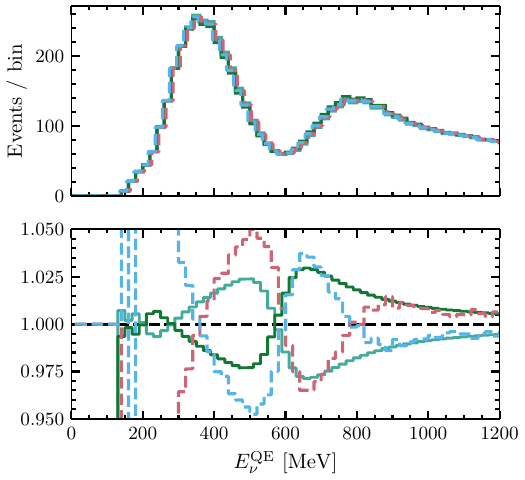}
    \caption{\footnotesize$E^{\mathrm{QE}}_{\nu}$, CC0$\pi$-H$_2$O, Hyper-K $\bar{\nu}_\mu$ osc.\ flux}
    \label{fig:Enuspect_anu_dm32_HK_reco}
  \end{subfigure}\\[8pt]
  \includegraphics[width=\linewidth]{Figures/legend_Fig1_HK_dCP.pdf}\\[2pt]
  \begin{subfigure}[b]{0.40\linewidth}
    \centering
    \includegraphics[width=\linewidth]{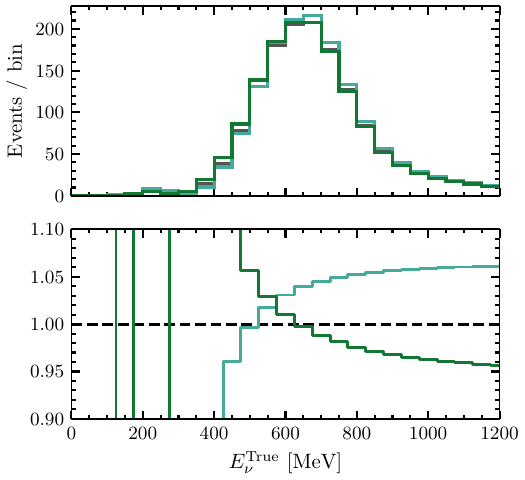}
    \caption{\footnotesize$E^{\mathrm{true}}_{\nu}$, CC0$\pi$-H$_2$O, Hyper-K $\bar{\nu}_e$ osc.\ flux}
    \label{fig:Enuspect_anu_dcp_HK_true}
  \end{subfigure}
  \begin{subfigure}[b]{0.40\linewidth}
    \centering
    \includegraphics[width=\linewidth]{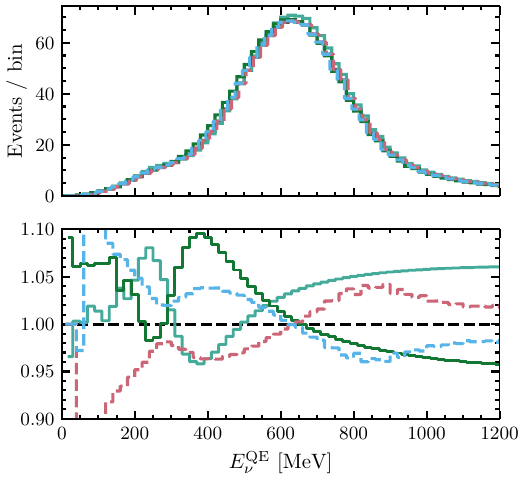}
    \caption{\footnotesize$E^{\mathrm{QE}}_{\nu}$, CC0$\pi$-H$_2$O, Hyper-K $\bar{\nu}_e$ osc.\ flux}
    \label{fig:Enuspect_anu_dcp_HK_reco}
  \end{subfigure}
  \caption{The NuWro simulated rate of CC0$\pi$ $\bar{\nu}_\mu$ or $\bar{\nu}_e$ interactions on water at the Hyper-K FD using the oscillated Hyper-K flux (see \autoref{tab:OscProbParams}) as a function of true ($E^{\mathrm{true}}_{\nu}$) or estimated ($E_{\nu}^{\mathrm{QE}}$) neutrino energy. For the former, variations of oscillation parameters corresponding to the Hyper-K ultimate target precision are shown. For the latter, these are compared to $E_{\nu}^{\mathrm{QE}}$ biased by $\pm$5~MeV.}
  \label{fig:Enuspect_anu_HK}
\end{figure}

\autoref{fig:Enuspect_anu_HK} shows that the impact of oscillation parameter variations and energy scale shifts is similar in shape and magnitude to the cases discussed in \autoref{fig:Enuspect_HK}. The energy scale variations for antineutrinos are slightly larger than a variation of 0.4\% on $\Delta m^2_{32}$ and retain the same shape as in the neutrino case. For $\bar\nu_e$ appearance, the energy scale variations are slightly smaller than the $\delta_{\mathrm{CP}}$ variation but remain comparable.

\begin{figure}[tbp]
\centering
  \includegraphics[width=\linewidth]{Figures/legend_Fig1_DUNE_dm32.pdf}\\[2pt]
  \begin{subfigure}[b]{0.40\linewidth}
    \centering
    \includegraphics[width=\linewidth]{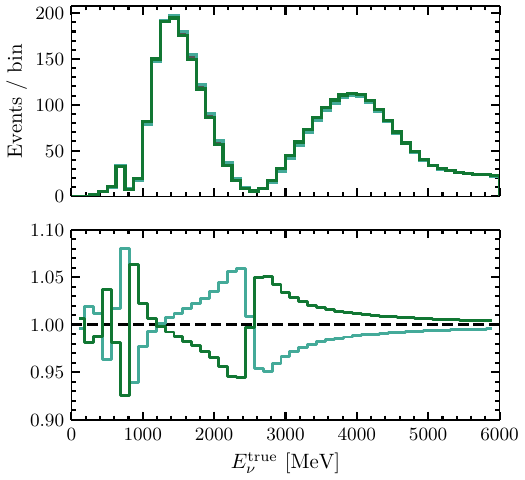}
    \caption{\footnotesize$E^{\mathrm{true}}_{\nu}$, CCINC-Ar, DUNE $\bar{\nu}_\mu$ osc.\ flux}
    \label{fig:Enuspect_anu_DUNE_dm32_true}
  \end{subfigure}
  \begin{subfigure}[b]{0.40\linewidth}
    \centering
    \includegraphics[width=\linewidth]{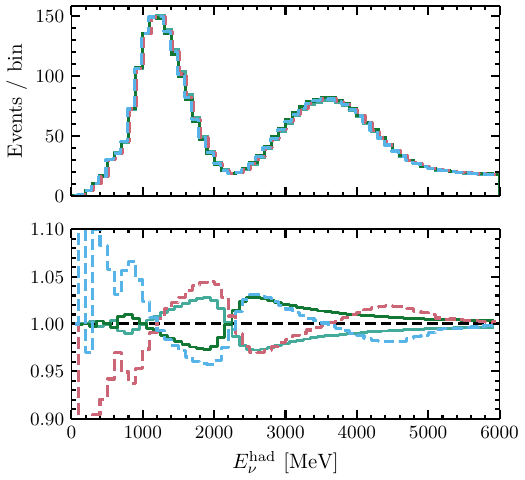}
    \caption{\footnotesize$E^{\mathrm{had}}_{\nu}$, CCINC-Ar, DUNE $\bar{\nu}_\mu$ osc.\ flux}
    \label{fig:Enuspect_anu_DUNE_dm32_reco}
  \end{subfigure}\\[8pt]
  \includegraphics[width=\linewidth]{Figures/legend_Fig1_DUNE_dCP.pdf}\\[2pt]
  \begin{subfigure}[b]{0.40\linewidth}
    \centering
    \includegraphics[width=\linewidth]{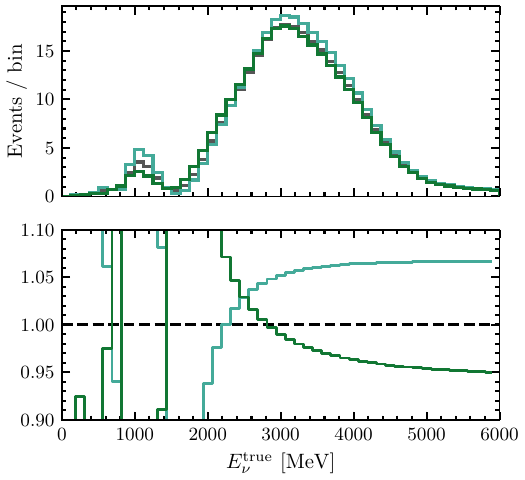}
    \caption{\footnotesize$E^{\mathrm{true}}_{\nu}$, CCINC-Ar, DUNE $\bar{\nu}_e$ osc.\ flux}
    \label{fig:Enuspect_anu_DUNE_dcp_true}
  \end{subfigure}
  \begin{subfigure}[b]{0.40\linewidth}
    \centering
    \includegraphics[width=\linewidth]{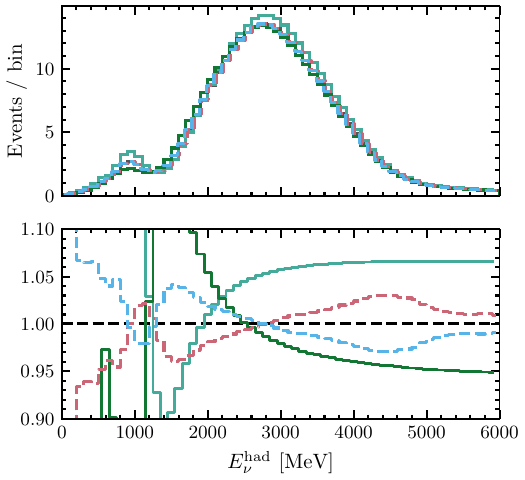}
    \caption{\footnotesize$E^{\mathrm{had}}_{\nu}$, CCINC-Ar, DUNE $\bar{\nu}_e$ osc.\ flux}
    \label{fig:Enuspect_anu_DUNE_dcp_reco}
  \end{subfigure}
  \caption{The NuWro simulated rate of CC inclusive $\bar{\nu}_\mu$ or $\bar{\nu}_e$ interactions on argon at the DUNE FD using the oscillated DUNE flux (see \autoref{tab:OscProbParams}) as a function of true ($E^{\mathrm{true}}_{\nu}$) or estimated ($E^{\mathrm{had}}_{\nu}$) neutrino energy. For the former, variations of oscillation parameters corresponding to the DUNE ultimate target precision are shown. For the latter, these are compared to $E^{\mathrm{had}}_{\nu}$ biased by $\pm$15~MeV.}
  \label{fig:Enuspect_anu_DUNE}
\end{figure}

\autoref{fig:Enuspect_anu_DUNE} shows the same variations as in \autoref{subsec:osc_spectra} applied to the DUNE antineutrino spectrum. For the $\bar\nu_\mu$ disappearance, the energy scale shift is comparable in both shape and magnitude to the shift in $\Delta m^2_{32}$, like in the neutrino case. However, the $\bar\nu_e$ appearance spectrum shows a significant difference with respect to the neutrino case. As discussed in \autoref{subsec:osc_spectra}, a shift in $\delta_{\mathrm{CP}}$ acts as a shape effect, with a response $\partial P/\partial\delta_{\mathrm{CP}} \propto -\sin(\Delta m^2_{32}L/4E + \delta_{\mathrm{CP}})$ that is, in vacuum, of similar size for $\nu_e$ and $\bar\nu_e$. What governs the visible distortion, however, is the size of this response \emph{relative} to the appearance probability $P(\nu_\mu\to\nu_e)$ itself. This probability is dominated by a $\delta_{\mathrm{CP}}$-independent term whose size is set by the mixing angles, $\propto \sin^2\theta_{23}\sin^2 2\theta_{13}$; the $\delta_{\mathrm{CP}}$-carrying interference term sits on top of it as a smaller modulation, suppressed by the ratio of the mass splittings $\alpha = \Delta m^2_{21}/\Delta m^2_{31} \approx 0.03$. Matter effects act on the dominant term with opposite sign for neutrinos and antineutrinos, enhancing it for $\nu_e$ and suppressing it for $\bar\nu_e$. A given $\delta_{\mathrm{CP}}$ shift therefore distorts the $\nu_e$ probability by a smaller relative amount and the $\bar\nu_e$ probability by a larger one. Consequently, for $\bar\nu_e$ the distortion from a $\delta_{\mathrm{CP}}$ variation is both larger than, and different in shape from, that produced by a 15~MeV energy-scale shift, whereas for $\nu_e$ the two are comparable. For Hyper-K, the much shorter baseline renders matter effects less pronounced, so this asymmetry between $\nu_e$ and $\bar\nu_e$ does not appear significant. 

\begin{figure}[htb]
\centering
  
  \includegraphics[width=\linewidth]{Figures/legend_Fig1_HK_dCP.pdf}\\[2pt]
  \begin{subfigure}[b]{0.40\linewidth}
    \centering
    \includegraphics[width=\linewidth]{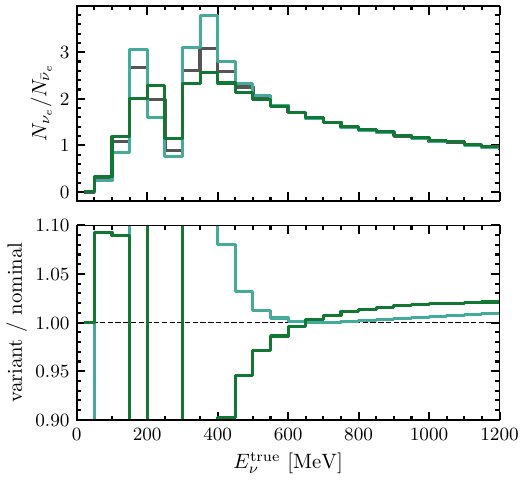}
    \caption{\footnotesize$E^{\mathrm{true}}_{\nu}$, CC0$\pi$-H$_2$O, Hyper-K $\nu_e/\bar\nu_e$ osc.\ flux}
    \label{fig:Enuspect_ratio_dcp_true}
  \end{subfigure}
  \begin{subfigure}[b]{0.40\linewidth}
    \centering
    \includegraphics[width=\linewidth]{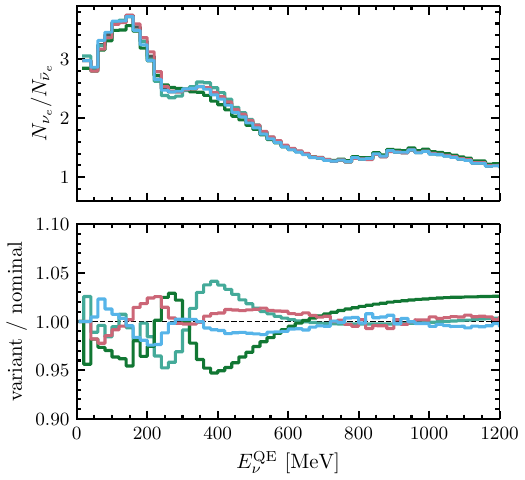}
    \caption{\footnotesize$E^{\mathrm{QE}}_\nu$, CC0$\pi$-H$_2$O, Hyper-K $\nu_e/\bar\nu_e$ osc.\ flux}
    \label{fig:Enuspect_ratio_dcp_reco}
  \end{subfigure}\\[8pt]
  \caption{The ratio of NuWro simulated rate of CC0$\pi$ $\nu_e$ to $\bar{\nu}_e$ interactions on water at the Hyper-K FD using the oscillated Hyper-K flux (see \autoref{tab:OscProbParams}) as a function of true ($E^{\mathrm{true}}_{\nu}$) or estimated ($E_{\nu}^{\mathrm{QE}}$) neutrino energy. For the former, variations of oscillation parameters corresponding to the Hyper-K ultimate target precision are shown. For the latter, these are compared to $E_{\nu}^{\mathrm{QE}}$ biased by $\pm$5~MeV.}
  \label{fig:Enuspect_ratio_HK}
\end{figure}

\begin{figure}[htb]
\centering
  
  \includegraphics[width=\linewidth]{Figures/legend_Fig1_DUNE_dCP.pdf}\\[2pt]
  \begin{subfigure}[b]{0.40\linewidth}
    \centering
    \includegraphics[width=\linewidth]{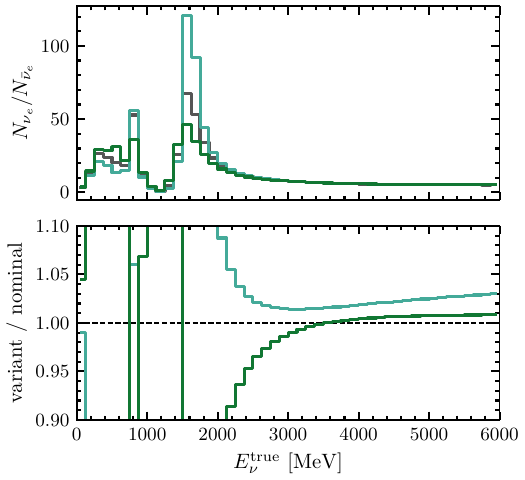}
    \caption{\footnotesize$E^{\mathrm{true}}_{\nu}$, CCINC-Ar, DUNE $\nu_e/\bar\nu_e$ osc.\ flux}
    \label{fig:Enuspect_DUNE_ratio_dcp_true}
  \end{subfigure}
  \begin{subfigure}[b]{0.40\linewidth}
    \centering
    \includegraphics[width=\linewidth]{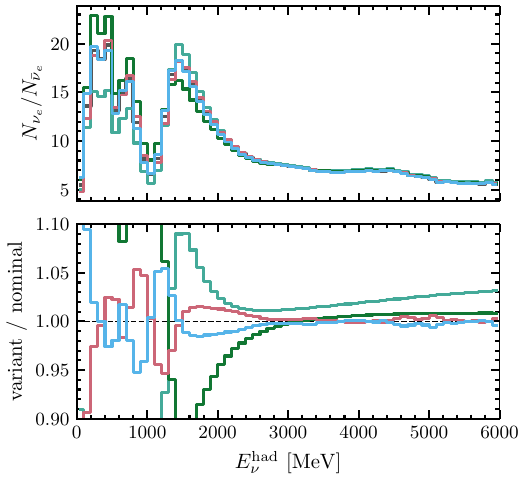}
    \caption{\footnotesize$E^{\mathrm{had}}_{\nu}$, CCINC-Ar, DUNE $\nu_e/\bar\nu_e$ osc.\ flux}
    \label{fig:Enuspect_DUNE_ratio_dcp_reco}
  \end{subfigure}\\[8pt]
  \caption{The ratio of NuWro simulated rate of CC inclusive $\nu_e$ to $\bar\nu_e$ interactions on argon at the DUNE FD using the oscillated DUNE flux (see \autoref{tab:OscProbParams}) as a function of true ($E^{\mathrm{true}}_{\nu}$) or estimated ($E_{\nu}^{\mathrm{had}}$) neutrino energy. For the former, variations of oscillation parameters corresponding to the DUNE ultimate target precision are shown. For the latter, these are compared to $E_{\nu}^{\mathrm{had}}$ biased by $\pm$15~MeV. }
  \label{fig:Enuspect_ratio_DUNE}
\end{figure}

In a real oscillation analysis, all samples are fitted simultaneously and so correlations between them, which the isolated single-sample spectra shown so far do not capture, are exploited to constrain the oscillation parameters. To investigate the impact of considering multiple oscillation samples in one simple way, we compare how energy scale shifts and $\delta_{\mathrm{CP}}$ change the ratio of the $\nu_e$ to $\bar\nu_e$ event rates\footnote{Note that the experiments plan to run with significantly different relative neutrino and antineutrino exposures (1:1 for DUNE and 1:3 for Hyper-K) and so a comparison of absolute value of the ratio between them is not meaningful}. \autoref{fig:Enuspect_ratio_HK} and \autoref{fig:Enuspect_ratio_DUNE} show that, for both the Hyper-K and DUNE cases, variations of $\delta_{\mathrm{CP}}$ around its nominal value of $-\pi/2$ distort this ratio in a way that differs in shape from, and is larger in magnitude than, the distortion induced by an energy-scale shift at the levels considered in this work. This demonstrates how the consideration of neutrino and antineutrino samples simultaneously can help reduce the impact of neutrino energy estimation bias on $\delta_{\mathrm{CP}}$ determination. We stress, however, that this does not imply that FSI or energy-scale effects are unimportant; it reflects only that, within the standard three-flavour PMNS framework and at the energy-scale precision considered here, their impact on the $\nu_e/\bar\nu_e$ ratio is subdominant to that of $\delta_{\mathrm{CP}}$. Searches for physics beyond this paradigm, such as non-standard interactions, may alter the $\nu_e$ and $\bar\nu_e$ spectra in ways that are degenerate with energy-scale changes, and so may require the neutrino energy scale to be controlled at a different, potentially more stringent, level than the one considered here.

\FloatBarrier

\section{Additional figures for FSI on/off variations}
\label{app:moreFSIvar}

\begin{figure}[htb]
\centering
  \includegraphics[width=\linewidth]{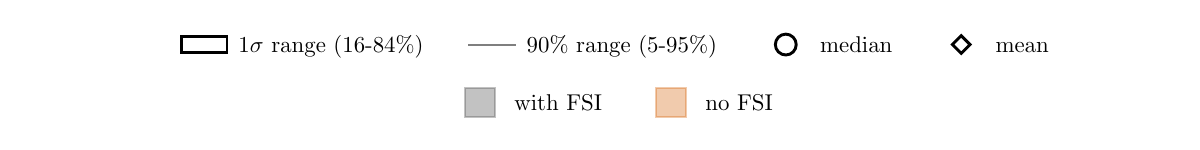}\\[4pt]
  \begin{subfigure}[b]{0.49\linewidth}
    \centering
    \includegraphics[width=\linewidth]{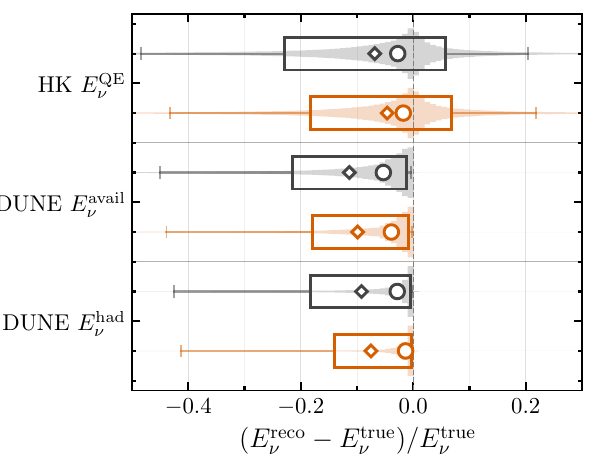}
    \caption{\footnotesize $\nu_\mu$}
  \end{subfigure}%
  \begin{subfigure}[b]{0.49\linewidth}
    \centering
    \includegraphics[width=\linewidth]{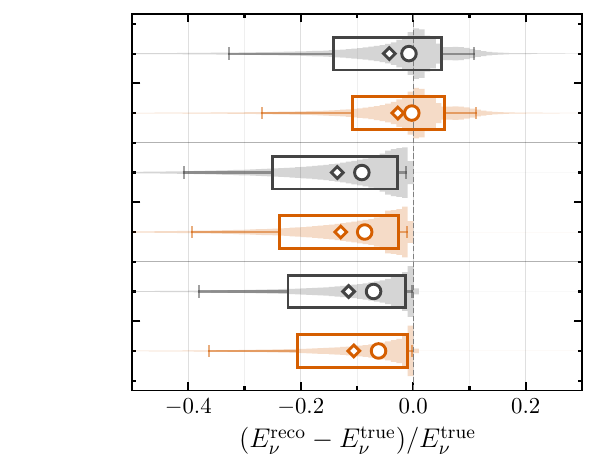}
    \caption{\footnotesize $\bar\nu_\mu$}
  \end{subfigure}
  \caption{The same as Fig.~\ref{fig:bw_piabs}, comparing the effect of turning on and off the INC within NuWro.}
  \label{fig:bw_fsi}
\end{figure}

\begin{figure}[htb]
\centering
  \includegraphics[width=\linewidth]{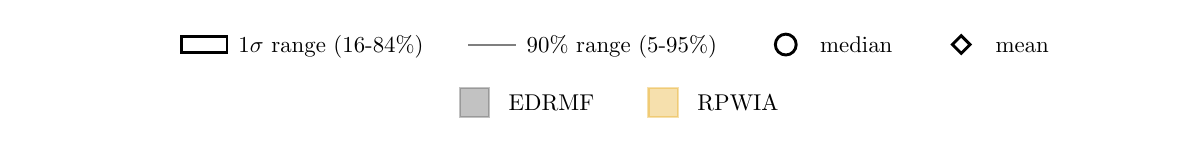}\\[4pt]
  \begin{subfigure}[b]{0.49\linewidth}
    \centering
    \includegraphics[width=\linewidth]{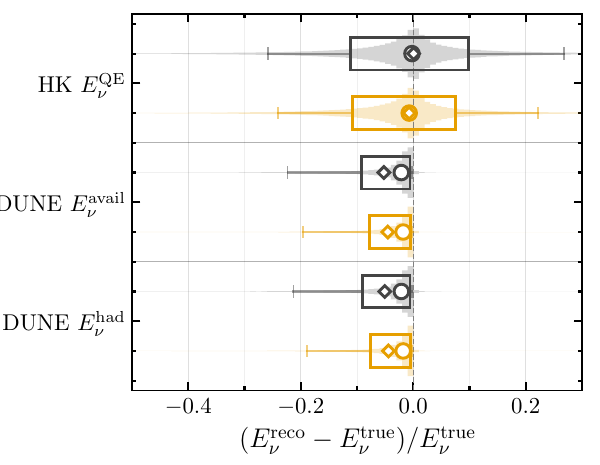}
    \caption{\footnotesize $\nu_\mu$}
  \end{subfigure}
  \caption{The same as Fig.~\ref{fig:bw_piabs}, comparing the NEUT ED-RMF and RPWIA nuclear models for CCQE interactions only.}
  \label{fig:bw_edrmf}
\end{figure}

\autoref{fig:bw_fsi} and \autoref{fig:bw_edrmf} summarise the impact of turning on and off the INC and the nuclear potential on the performance of the neutrino energy estimators considered in this work, supplementing the specific FSI variations considered in \autoref{fig:bw_piabs}, \autoref{fig:bw_mfp}  and \autoref{fig:bw_genie}.

\FloatBarrier

\section{Relative bias studies}
\label{app:relBias}

In \autoref{tab:FSIVar} the sensitivity to the energy-estimation bias to FSI model variations is quantified in \textit{absolute} terms (as the difference between the estimated and true neutrino energy in MeV). Since Hyper-K and DUNE operate at substantially different mean neutrino energies, it is also informative to express the bias \textit{relative} to the true neutrino energy, $(E^{\mathrm{reco}}_{\nu}-E^{\mathrm{true}}_{\nu})/E^{\mathrm{true}}_{\nu}$, which places the two experiments on a more common footing. \autoref{tab:FSIVar_combined} reproduces the absolute shifts of \autoref{tab:FSIVar} whilst also reporting their relative counterparts. We recall that the $\sim$5~MeV (Hyper-K) and $\sim$15~MeV (DUNE) absolute targets broadly correspond to a shift of $\sim$0.5\% of the respective mean neutrino energy (see \autoref{subsec:osc_spectra}); the $0.5\%$ threshold used to flag the relative shifts is therefore the common relative counterpart of these two absolute requirements. We note, however, that the absolute and relative measures are not equivalent: normalising by $E^{\mathrm{true}}_{\nu}$ gives proportionally less weight to high-energy interactions, where the absolute bias is largest, so the two measures provide a slightly different quantification of the change in energy estimation bias.

Overall, \autoref{tab:FSIVar_combined} shows that the main conclusions of \autoref{sec:results} are preserved when considering relative bias shifts, although several details differ in instructive ways. For the DUNE case, varying the full INC model remains the dominant effect. The pion-absorption variation, which reaches the 15~MeV scale for $E^{\mathrm{had}}_{\nu}$ in absolute terms, becomes a much smaller effect in relative terms. The relative sensitivity to the nucleon mean free path (MFP) is also reduced but remains significant for antineutrinos, where it skews the bias distribution. For the Hyper-K case, the dominant effects are the pion absorption probability and FSI beyond the semi-classical INC. In particular, the relative shift in the mean bias between the ED-RMF and RPWIA models now exceeds the $0.5\%$ threshold, whereas the corresponding absolute shift lies below the 5~MeV threshold. This further highlights the importance of FSI modelling beyond the cascade picture for Hyper-K. A comparable relative shift from the ED-RMF/RPWIA comparison is also seen for DUNE. We recall, however, that it is computed for CCQE interactions only, which make up less than one third of the DUNE event rate, so this particular conclusion is harder to generalise.

\begin{table}[htb]
\centering
\scriptsize
\setlength{\tabcolsep}{4pt}
\renewcommand{\arraystretch}{1.15}
\begin{tabular}{@{}c l r r r r r r r r r r@{}}
\toprule
\textbf{Flavour} & \textbf{Observable} & \multicolumn{2}{c}{\textbf{FSI / no FSI}} & \multicolumn{2}{c}{\textbf{$\pi_{\mathrm{abs}}\pm31\%$}} & \multicolumn{2}{c}{\textbf{NN MFP $\pm30\%$}} & \multicolumn{2}{c}{\textbf{INC model var.}} & \multicolumn{2}{c}{\textbf{Nuc. Pot. on/off}} \\
\cmidrule(lr){3-4}\cmidrule(lr){5-6}\cmidrule(lr){7-8}\cmidrule(lr){9-10}\cmidrule(lr){11-12}
 &  & median & mean & median & mean & median & mean & median & mean & median & mean \\
\midrule
\multicolumn{12}{c}{\textbf{Absolute bias [MeV]}} \\
\midrule
\multirow{3}{*}{\rotatebox[origin=c]{90}{\textbf{$\nu_\mu$}}} & Hyper-K $E_{\nu}^{\mathrm{QE}}$ & \cellcolor{red!12}10.1 & \cellcolor{red!12}35.8 & \cellcolor{red!12}5.8 & \cellcolor{red!12}16.4 & 0.2 & 1.0 & 0.8 & \cellcolor{red!12}9.4 & \cellcolor{red!12}7.0 & 4.4 \\
 & DUNE $E_{\nu}^{\mathrm{avail}}$ & \cellcolor{red!12}28.7 & \cellcolor{red!12}79.4 & 1.6 & 2.9 & 10.6 & 4.2 & \cellcolor{red!12}44.5 & \cellcolor{red!12}32.2 & 0.3 & 4.5 \\
 & DUNE $E_{\nu}^{\mathrm{had}}$ & \cellcolor{red!12}88.7 & \cellcolor{red!12}81.9 & 13.8 & 10.7 & \cellcolor{red!12}16.3 & 2.4 & \cellcolor{red!12}39.2 & \cellcolor{red!12}27.1 & 0.3 & 4.1 \\
\midrule
\multirow{3}{*}{\rotatebox[origin=c]{90}{\textbf{$\bar\nu_\mu$}}} & Hyper-K $E_{\nu}^{\mathrm{QE}}$ & \cellcolor{red!12}7.0 & \cellcolor{red!12}25.6 & 3.6 & \cellcolor{red!12}12.5 & 0.3 & 1.1 & 0.1 & 2.0 &  &  \\
 & DUNE $E_{\nu}^{\mathrm{avail}}$ & \cellcolor{red!12}39.8 & \cellcolor{red!12}35.4 & 7.7 & 6.5 & 14.7 & \cellcolor{red!12}16.2 & \cellcolor{red!12}39.8 & \cellcolor{red!12}16.4 &  &  \\
 & DUNE $E_{\nu}^{\mathrm{had}}$ & \cellcolor{red!12}55.1 & \cellcolor{red!12}43.0 & \cellcolor{red!12}17.8 & 13.7 & \cellcolor{red!12}17.3 & \cellcolor{red!12}18.7 & \cellcolor{red!12}61.7 & \cellcolor{red!12}37.0 &  &  \\
\midrule
\multicolumn{12}{c}{\textbf{Relative bias [\%]}} \\
\midrule
\multirow{3}{*}{\rotatebox[origin=c]{90}{\textbf{$\nu_\mu$}}} & Hyper-K $E_{\nu}^{\mathrm{QE}}$ & \cellcolor{red!12}1.0 & \cellcolor{red!12}2.2 & \cellcolor{red!12}0.6 & \cellcolor{red!12}0.9 & 0.0 & 0.1 & 0.0 & 0.3 & \cellcolor{red!12}0.5 & \cellcolor{red!12}0.8 \\
 & DUNE $E_{\nu}^{\mathrm{avail}}$ & \cellcolor{red!12}1.4 & \cellcolor{red!12}1.5 & 0.0 & 0.0 & 0.4 & 0.4 & \cellcolor{red!12}1.2 & \cellcolor{red!12}1.0 & 0.3 & \cellcolor{red!12}0.7 \\
 & DUNE $E_{\nu}^{\mathrm{had}}$ & \cellcolor{red!12}1.5 & \cellcolor{red!12}1.7 & 0.1 & 0.2 & 0.4 & 0.3 & \cellcolor{red!12}1.9 & \cellcolor{red!12}1.4 & 0.3 & \cellcolor{red!12}0.7 \\
\midrule
\multirow{3}{*}{\rotatebox[origin=c]{90}{\textbf{$\bar\nu_\mu$}}} & Hyper-K $E_{\nu}^{\mathrm{QE}}$ & \cellcolor{red!12}0.5 & \cellcolor{red!12}1.5 & 0.3 & \cellcolor{red!12}0.7 & 0.0 & 0.0 & 0.1 & 0.1 &  &  \\
 & DUNE $E_{\nu}^{\mathrm{avail}}$ & \cellcolor{red!12}0.5 & \cellcolor{red!12}0.6 & 0.1 & 0.2 & 0.4 & 0.4 & \cellcolor{red!12}1.1 & \cellcolor{red!12}1.1 &  &  \\
 & DUNE $E_{\nu}^{\mathrm{had}}$ & \cellcolor{red!12}0.9 & \cellcolor{red!12}0.9 & 0.3 & 0.3 & 0.4 & \cellcolor{red!12}0.5 & \cellcolor{red!12}1.6 & \cellcolor{red!12}1.5 &  &  \\
\bottomrule
\end{tabular}
\normalsize
\caption{The shift in the mean and median neutrino energy estimation bias due to different FSI variations for the Hyper-K and DUNE neutrino and antineutrino cases. The top half reports the absolute shift in MeV, and the bottom half the same shift expressed as a fraction of $E^{\mathrm{true}}_{\nu}$ (in percent). The numbers reported for the INC model variation are derived from the two INC models that give the largest spread in the mean. Red boxes indicate that the variation is larger than 5\,MeV / 15\,MeV (absolute, Hyper-K and DUNE respectively) or $0.5\%$ (relative), which is broadly indicative of how well the neutrino energy reconstruction scale must be controlled (see \autoref{sec:enurec}). ``Nuc. Pot.'' stands for the nuclear potential considered in \autoref{subsec:beyondcasc}, for which the table reports a shift derived considering only CCQE interactions.}
\label{tab:FSIVar_combined}
\end{table}
\FloatBarrier

\bibliography{biblo.bib}
\bibliographystyle{unsrt.bst}

\end{document}